\documentclass[a4paper,fleqn,usenatbib]{mnras}
\usepackage{multirow}
\usepackage{booktabs}
\usepackage{amstext}
\usepackage{graphicx,color}
\usepackage{amssymb}
\usepackage{hyperref}
\usepackage{amsmath}
\usepackage{epstopdf}
\usepackage{comment}
\usepackage{gensymb}
\usepackage{textcase}
\def\lowc{\expandafter\lowercaseA\expandafter{\iffalse}\fi}
\long\def\lowercaseA#1{\lowercase{#1}\egroup}
\usepackage{newtxtext,newtxmath}
\usepackage{comment,color}
\usepackage[T1]{fontenc}
\usepackage{ae,aecompl}

\title[]{Can astrophysical neutrinos trace the origin of the detected ultra-high energy cosmic rays?}

\author[A. Palladino et al.]{
Andrea Palladino,$^{1}$\thanks{E-mail: andrea.palladino@desy.de}
Arjen van Vliet,$^{1}$
Walter Winter$^{1}$
and
Anna Franckowiak$^{1}$
\\
$^{1}$DESY, Platanenallee 6, 15738 Zeuthen, Germany
}

\date{Accepted XXX. Received YYY; in original form ZZZ}

\pubyear{2020}

\begin{document}
\label{firstpage}
\pagerange{\pageref{firstpage}--\pageref{lastpage}}
\maketitle

\begin{abstract}
Since astrophysical neutrinos are produced in the interactions of cosmic rays, identifying the origin of cosmic rays using directional correlations with neutrinos is one of the most interesting possibilities of the field. For that purpose, especially the Ultra-High Energy Cosmic Rays (UHECRs) are promising, as they are deflected less by extragalactic and Galactic magnetic fields than cosmic rays at lower energies. However, photo-hadronic interactions of the UHECRs limit their horizon, while neutrinos do not interact over cosmological distances. We study the possibility to search for anisotropies by investigating neutrino-UHECR correlations from the theoretical perspective, taking into account the UHECR horizon, magnetic-field deflections, and the cosmological source evolution. Under the assumption that the neutrinos and UHECRs all come from the same source class, we demonstrate that the non-observation of neutrino multiplets strongly constrains the possibility to find neutrino-UHECR correlations. 
\end{abstract}


\section{Introduction}

The first hints for the sources of Ultra-High Energy Cosmic Rays (UHECRs) have been observed, as correlations with catalogs of starburst galaxies and Active Galactic Nuclei (AGN) reach the $\sim$4$\sigma$ level \citep{Aab:2018chp, Caccianiga:2019hlc}. However, a definitive answer to the question of the origin of UHECRs remains illusive. Moreover, the discovery of a diffuse flux of high energy neutrinos by IceCube in 2013 \citep{Aartsen:2013jdh} has opened new possibilities to search for the sources of UHECRs. Indeed, it is commonly believed that sources that are able to accelerate protons up to very high energies are also good neutrino-emitter candidates. The knowledge of the cosmic-ray flux and the connection expected between these two astrophysical messengers have already been discussed more than 50 years ago \citep{Beresinsky:1969qj}. In the work presented here we investigate the probability to observe spatial correlations between high energy neutrinos and cosmic rays, assuming that they are produced by the same source class. 

Three different experimental searches for a correlation of high energy cosmic rays and neutrinos have been performed \citep{Aartsen:2015dml,Schumacher:2019qdx}. All three analyses use an UHECR sample consisting of events above 52\,EeV and 57\,EeV recorded by the Pierre Auger Observatory and Telescope Array (TA) experiments, respectively. First, a cross-correlation analysis has counted the number of neutrino-UHECR pairs separated by less than a given angular distance. This number has been compared to the simulated number of random pairs within the same angular distance. This analysis uses a high-purity but low-statistic sample of IceCube events, focused on energies above 60~TeV. Second, using the same neutrino sample as the first analysis, a stacking of the neutrino arrival directions has been applied, searching for coincident sources of cosmic rays. The cosmic-ray arrival direction has been smeared by the observatories' angular uncertainties and an energy-dependent Galactic magnetic deflection based on a Galactic Magnetic Field (GMF) model. The third analysis has used a high-statistics, but low-purity neutrino data sample of both IceCube and ANTARES events in the energy range from 1\,TeV to 1\,PeV, which was optimized for the search of neutrino point sources. A search for neutrino point sources in the vicinity of the UHECR arrival directions has been performed using a stacking analysis. This is essentially a point-source stacking analysis with a spatial prior given by the UHECR's direction uncertainty, smeared with an energy-dependent Galactic magnetic deflection. No significant excess has been found in any of the three analyses. These non-detections of correlations suggest to analyze the problem from a theoretical point of view, in order to understand if there are conditions in which correlations are likely to be observed.

It is important to recall that neutrinos and cosmic rays propagate in a very different manner. While this is clear immediately from the basic properties of the particles, it will turn out to be one of the main reasons which will prevent us from observing correlations -- and which we are going to quantify in this work.
While neutrinos propagate on geodesics, cosmic rays are strongly deflected by both extragalactic and Galactic magnetic fields. Moreover, astrophysical neutrinos lose energy only due to the adiabatic expansion of the Universe, while UHECRs, depending on energy and composition, lose energy due to adiabatic expansion, Bethe Heitler pair production, photo-meson production and photo-disintegration on cosmic background light. As a consequence, UHECRs can only reach the Earth when they are produced in the local Universe. On the contrary, astrophysical neutrinos can reach the Earth over cosmological distances -- which  implies that the source evolution of the source class plays a crucial role in our study. We will use three different examples characterizing different classes of sources: negative source evolution (which may characterize Tidal Disruption Events (TDEs) \citep{Sun:2015bda} or low-luminosity blazars~\citep{Ajello:2013lka}), flat evolution (which is the simplest assumption of a source class present independent of redshift), and Star Formation Rate (SFR) evolution~\citep{Yuksel:2008cu} (which roughly describes most conventional source classes, including normal galaxies, starburst galaxies, high luminosity BL Lacs, Flat Spectrum Radio Quasars (FSRQs) and Gamma-Ray Bursts (GRBs)).

In this study, we scrutinize the hypothesis that a common origin of the astrophysical neutrinos and UHECRs can be identified from directional correlations, assuming that both UHECRs and neutrinos stem from the same source class. We will define a model using a Monte Carlo simulation, based on simple counting statistics, and we will quantify the impact of magnetic-field deflections, cosmic source evolution and particle interactions during propagation. Using the same methods, we will also point out that the observation of neutrino-UHECR correlations implies the observation of neutrino multiplets. With the term multiplet we mean two or more tracks above 200 TeV (the threshold of the throughgoing muon sample) coming from the same direction, within the typical angular resolution of track-like events (i.e. $1^\circ$) and within the lifetime of the IceCube experiment.\footnote{In principle, it is also possible to use a lower energy threshold, but in this case it is hard to discriminate the signal from the background. Moreover, neutrino multiplets have already been observed below $\sim$100 TeV, as in the neutrino flare from TXS 0506+056 during 2014-15 \citep{IceCube:2018cha}.} Our purpose is to identify the remaining parameter space where a common neutrino-UHECR origin can be identified.

\section{Method}
\label{sec:method}

In a nutshell, we use a Monte Carlo simulation to extract a number of sources randomly distributed in the sky, following a given redshift distribution, assuming that these sources are the common sources of neutrinos and UHECRs. Then we propagate both neutrinos and cosmic rays, taking into account the energy loss of neutrinos and cosmic rays as well as the deflection of cosmic rays.
Our simulated rates are chosen to match observations provided by Auger and IceCube by taking into account the number of events detected by the two experiments. Finally, we compute the probability to discriminate a $5 \sigma$ signal (i.e.\ cosmic-ray events within a certain angular distance from neutrino events) from a background consisting of isotropically-distributed events.

\subsection{Neutrinos}

\begin{figure*}
\centering
\includegraphics[width=0.4\textwidth]{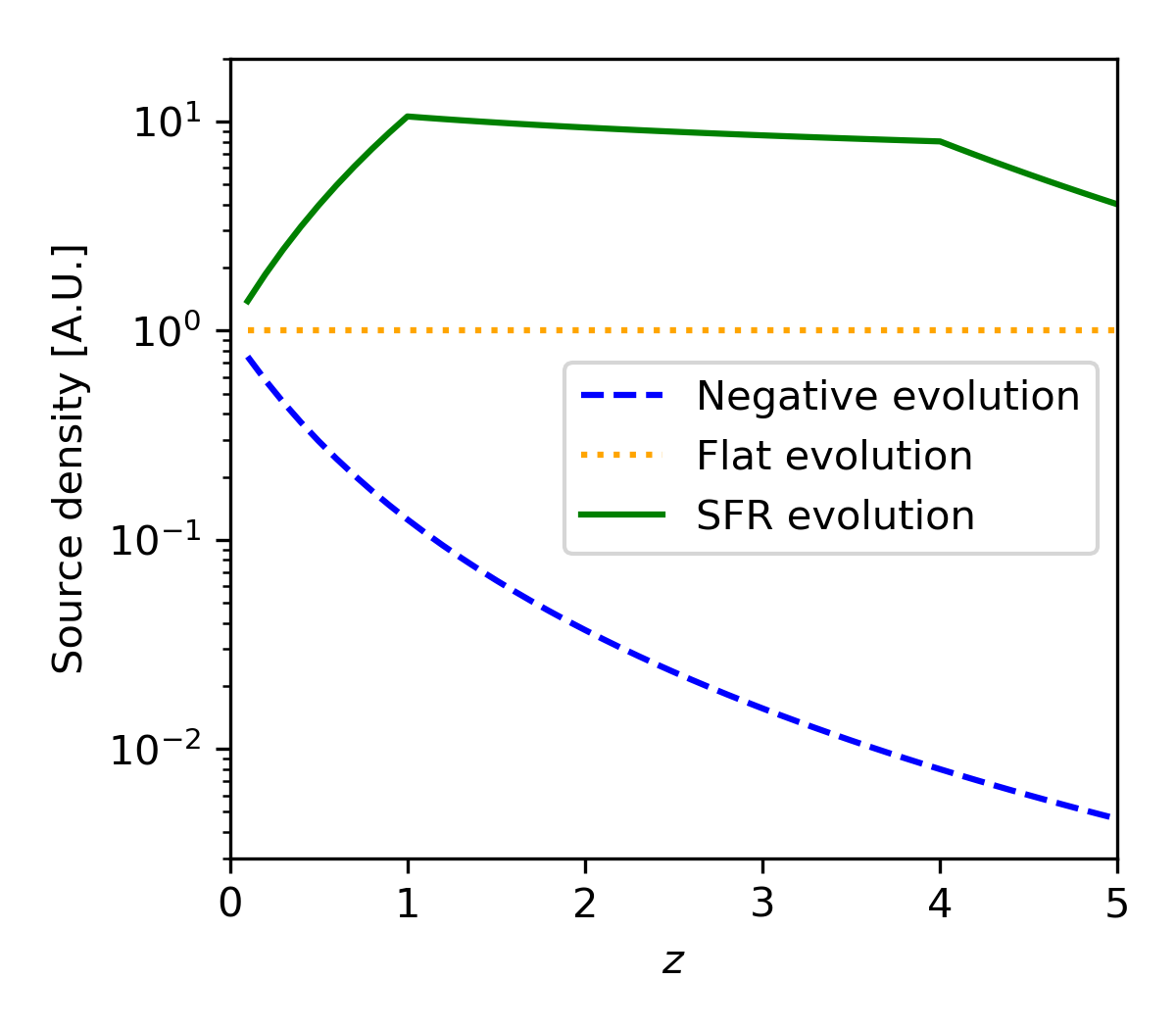}
\includegraphics[width=0.4\textwidth]{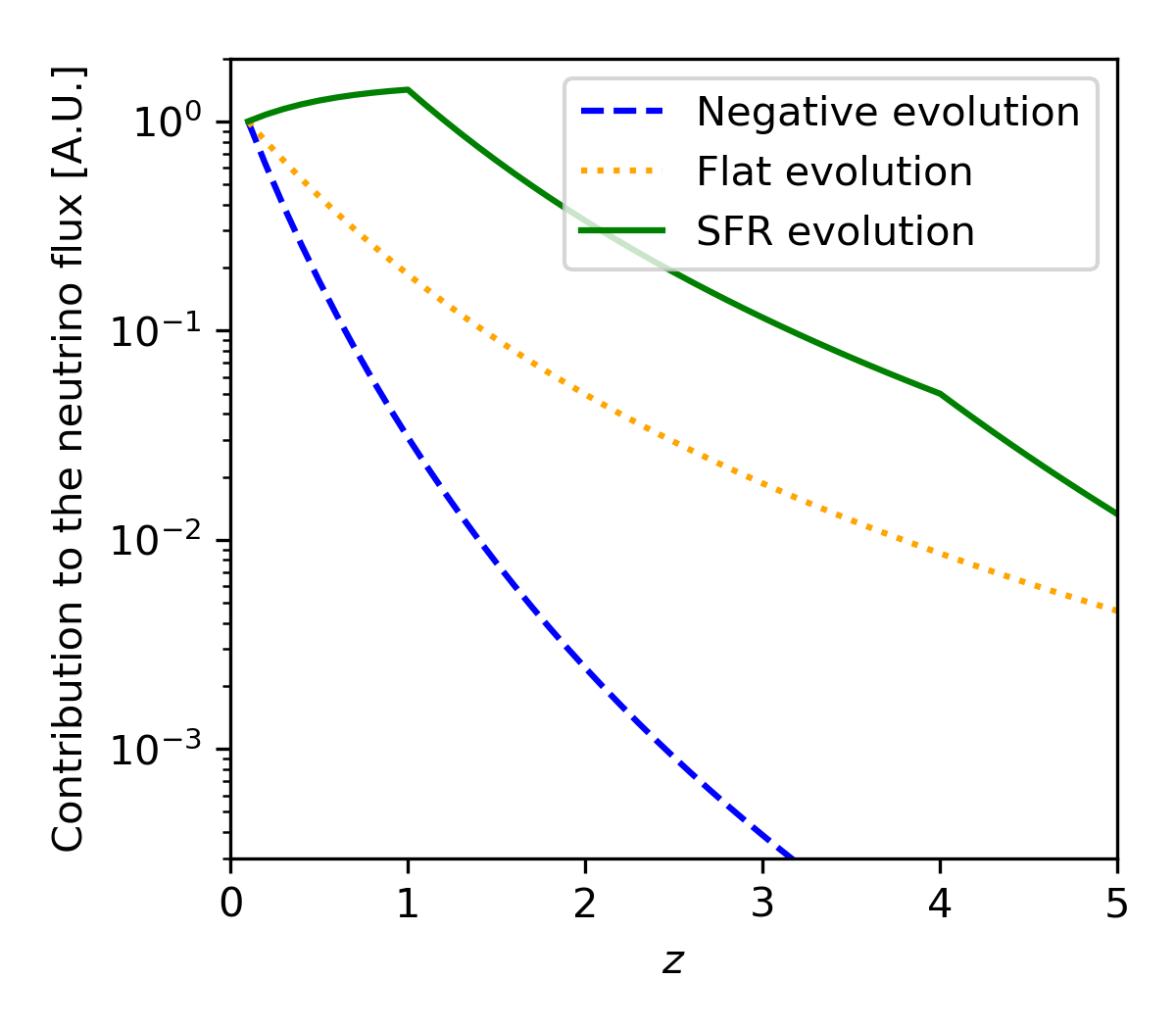} \\
\includegraphics[width=0.4\textwidth]{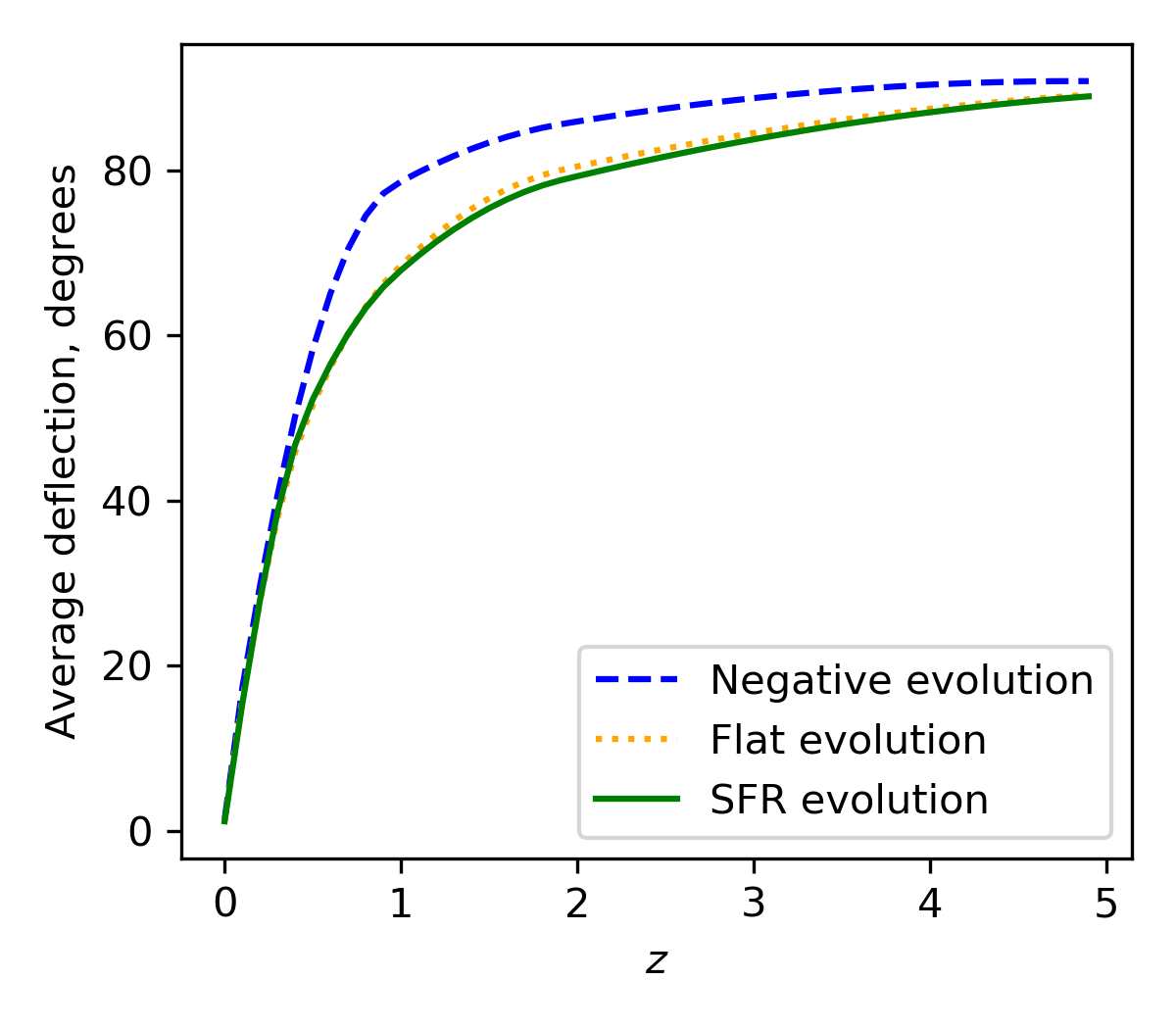}
\includegraphics[width=0.4\textwidth]{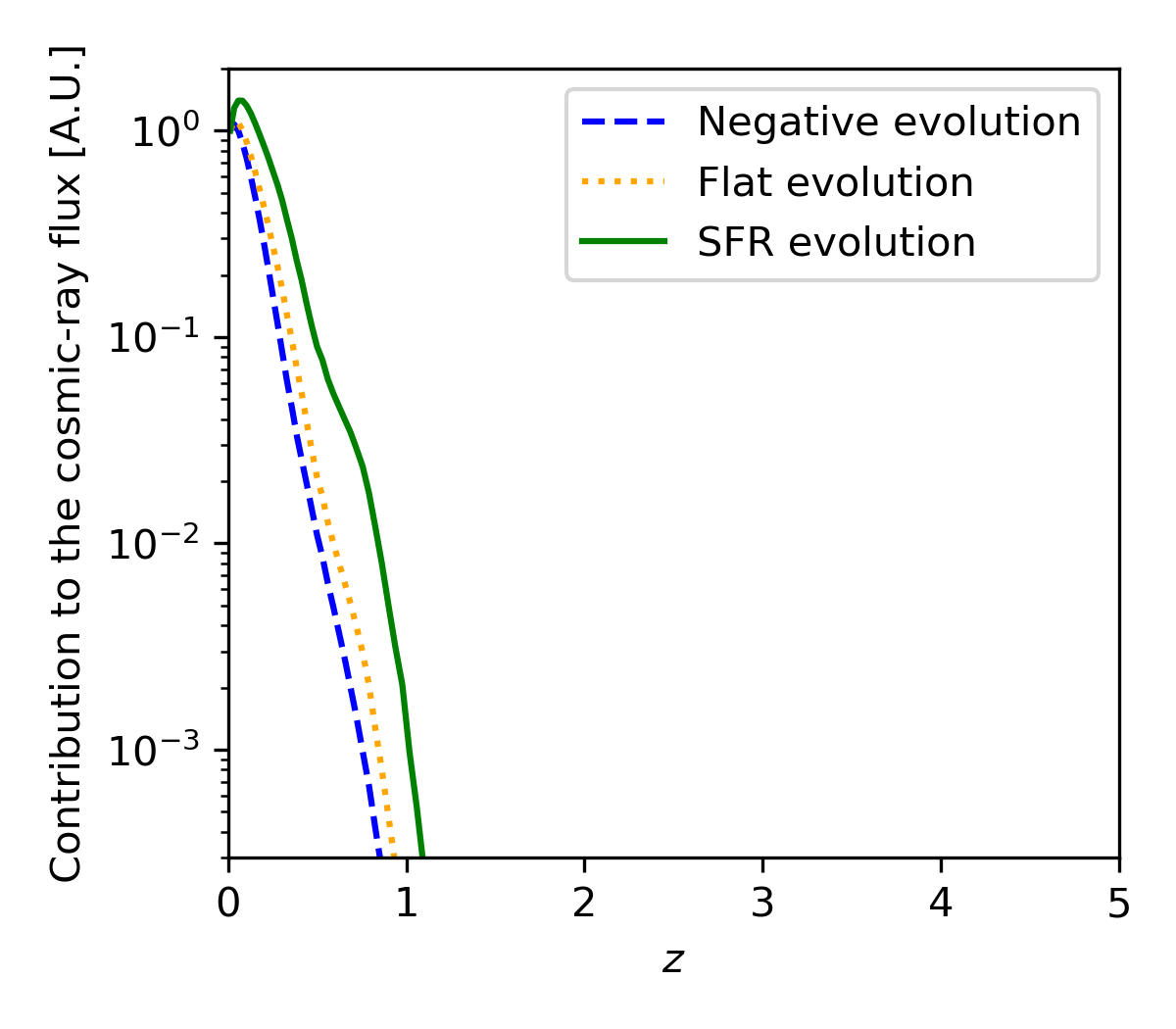}
\caption{Upper left panel: the source density as a function of redshift, for three different cases: negative evolution (blue, dashed), flat evolution (orange, dotted) and sources following  the SFR (green, solid). Upper right panel: the corresponding contribution to the neutrino flux as a function of redshift for the three different source evolution scenarios, considering a power-law spectrum. Lower left panel: the median deflection of cosmic rays for all UHECRs with $E_{\text{CR}}>10^{18.5}$~eV, as a function of redshift. Lower right panel: the corresponding contribution to the cosmic-ray flux as a function of redshift.}
\label{plot:nuevo}
\end{figure*}

First of all, we define three different source evolutions ($\rho(z)$), as a function of redshift ($z$): \textit{i)} a negative source evolution, following an $(1+z)^{-3}$ behavior as a function of redshift; \textit{ii)} a flat source evolution; \textit{iii)} SFR evolution. These source evolutions are shown in the upper left panel of Fig.~\ref{plot:nuevo}. Other details concerning the source evolutions and the definition of the required cosmological functions are reported in Appendix~\ref{sec:sevo}. In the upper right panel of Fig.~\ref{plot:nuevo} we show the contribution to the neutrino luminosity as a function of redshift. This function $f_\nu(z)$  is proportional to
\begin{equation}
f_\nu(z) \propto \rho(z) \times \frac{\text{d}V_c}{\text{d}z} \times D_\ell(z)^{-2} \, ,
\label{eq:probnu}
\end{equation}
where $\frac{\text{d}V_c}{\text{d}z}$ is the comoving volume and $D_\ell(z)$ is the luminosity distance\footnote{Note that in that equation it is implicitly assumed that the neutrino flux is characterized by a power-law spectrum, at least in the energy region in which the detector is sensitive. The consequences of a different scenario are discussed in Sec~\ref{sec:discussion}.} (both functions are defined in Appendix~\ref{sec:sevo}). We assume that sources are standard candles, which means that no source-luminosity dependency is contained in the previous expression. Let us recall that neutrinos are not deflected during the propagation and they are only affected by adiabatic energy losses. 

\subsection{Cosmic rays}

The case of UHECRs is more complex than the neutrino case. Due to the additional energy-loss processes on the way, very energetic cosmic rays can only come from the local Universe, while neutrinos can reach the Earth from distant sources as well. In addition, cosmic rays are deflected during the propagation by both extragalactic and Galactic magnetic fields.

To take into account these effects we compute the spectrum and composition of UHECRs using CRPropa~3 \citep{Batista:2016yrx} in the one-dimensional (1D) mode. For each source-evolution scenario values for the spectral index ($\gamma$), maximum rigidity ($R_{\text{max}}$) and composition (specified by proton ($f_\text{p}$), helium ($f_\text{He}$), nitrogen ($f_\text{N}$) and silicon ($f_\text{Si}$) fractions) at the sources are obtained that provide good fits to both the measured spectrum and composition of UHECRs by Auger, assuming an injected source spectrum of 
\begingroup
\small
\thinmuskip=\muexpr\thinmuskip*5/8\relax
\medmuskip=\muexpr\medmuskip*5/8\relax
$$
	\frac{\text{d}N_i}{\text{d}E} \propto
	\begin{cases}	
		\; f_i  E^{-\gamma} & \text{ for } E < Z_i R_\text{max} \, , \\ 
		\; f_i E^{-\gamma} \exp\left(1  - \frac{E}{Z_i R_\text{max}} \right) & \text{ for } E \geq Z_i R_\text{max} \, , 
	\end{cases}
$$
\endgroup
with $E$ and $Z_i$ the cosmic-ray energy and charge at the source, respectively. For the negative source evolution case the best-fit values were taken from \citet{Aab:2016zth}, for the flat and SFR source evolution they were taken from \citet{AlvesBatista:2018zui}. In all three cases the same UHECR data set as measured by Auger was used in the fits, for details see \citet{Aab:2016zth}. These three fits all assumed EPOS LHC as hadronic interaction model in UHECR air showers and the full $X_{\text{max}}$ distributions, including their associated errors, were fitted for each energy bin. See Table~\ref{tab:CRbestfitparam} for the best-fit values that were found in \citet{Aab:2016zth} and \citet{AlvesBatista:2018zui} and are used in this work.

\begin{table}
\centering
\caption{Best-fit parameters, obtained from \citet{Aab:2016zth} and \citet{AlvesBatista:2018zui}, used in this work for the UHECR simulations.}
  \begin{tabular}{l c c c c c c c}
  \hline
     $\rho(z)$ & $\gamma$ & $R_\text{max}/\text{V}$ & $f_\text{p}$ & $f_\text{He}$ & $f_\text{N}$ & $f_\text{Si}$ \\ 
   \hline
     Neg. & $1.42$ & $10^{18.85}$ & $0.07$ & $0.34$ & $0.53$ & $0.06$ \\  
     Flat & $-1.0$ & $10^{18.2}$ & $0.6726$ & $0.3135$ & $0.0133$ & $0.0006$ \\ 
     SFR & $-1.3$ & $10^{18.2}$ & $0.1628$ & $0.8046$ & $0.0309$ & $0.0018$ \\ 
  \hline
  \end{tabular}
  \label{tab:CRbestfitparam}
\end{table}

For these three scenarios the contribution to the UHECR flux as a function of the redshift has been obtained for cosmic-ray energies at Earth of $E_{\text{CR}}>10^{18.5}$~eV, see Fig.~\ref{plot:nuevo} lower right panel. Due to the different energy-loss processes the UHECRs can only arrive from relatively nearby sources. The redshift dependence of the contribution to the UHECR flux is, therefore, quite similar for the three different source-evolution scenarios considered here, and is very different from the neutrino case (Fig.~\ref{plot:nuevo} upper right panel). 

The expected deflections of cosmic rays due to Extragalactic Magnetic Fields (EGMFs) ($\Delta_{\text{EGMF}} (z)$) as a function of the redshift for $E_{\text{CR}}>10^{18.5}$~eV have been computed in a 3D simulation with CRPropa in a structured EGMF of \citet{Hackstein:2017pex}, implementing the same best-fit parameters for the three source-evolution scenarios (Table~\ref{tab:CRbestfitparam}) as well as a continuous distribution of identical sources. From the six EGMF models described in \citet{Hackstein:2017pex}, we use the \lq\lq astrophysical model\rq\rq . This model has the smallest filling factors (fraction of the total space filled with magnetic fields of a certain strength or stronger) for the strongest magnetic fields of the six different modes. This means that it can be expected to give the smallest average deflection, leading to a rather conservative estimate for the deflections of UHECRs in EGMFs. Reflective boundary conditions were implemented in the simulations to properly include sources at large distances. Adiabatic energy losses have not been included in these simulations as, in 3D simulations, the total travel time (redshift) of the particle is not known at the start of the simulation due to the deflections in EGMFs. The inclusion of adiabatic energy losses would, however, only have a small effect on $\Delta_{\text{EGMF}} (z)$ for large redshifts, which are not relevant for the expected neutrino-UHECR correlations. The resulting average deflections as a function of redshift ($\langle \Delta_{\text{EGMF}} \rangle (z)$) are given in Fig.~\ref{plot:nuevo} lower left panel. Here $\langle \Delta_{\text{EGMF}} \rangle (z)$ is largest in the negative evolution scenario due to the heavier composition at the sources in that scenario.

\begin{figure*}
	\centering
	\includegraphics[width=0.4\textwidth]{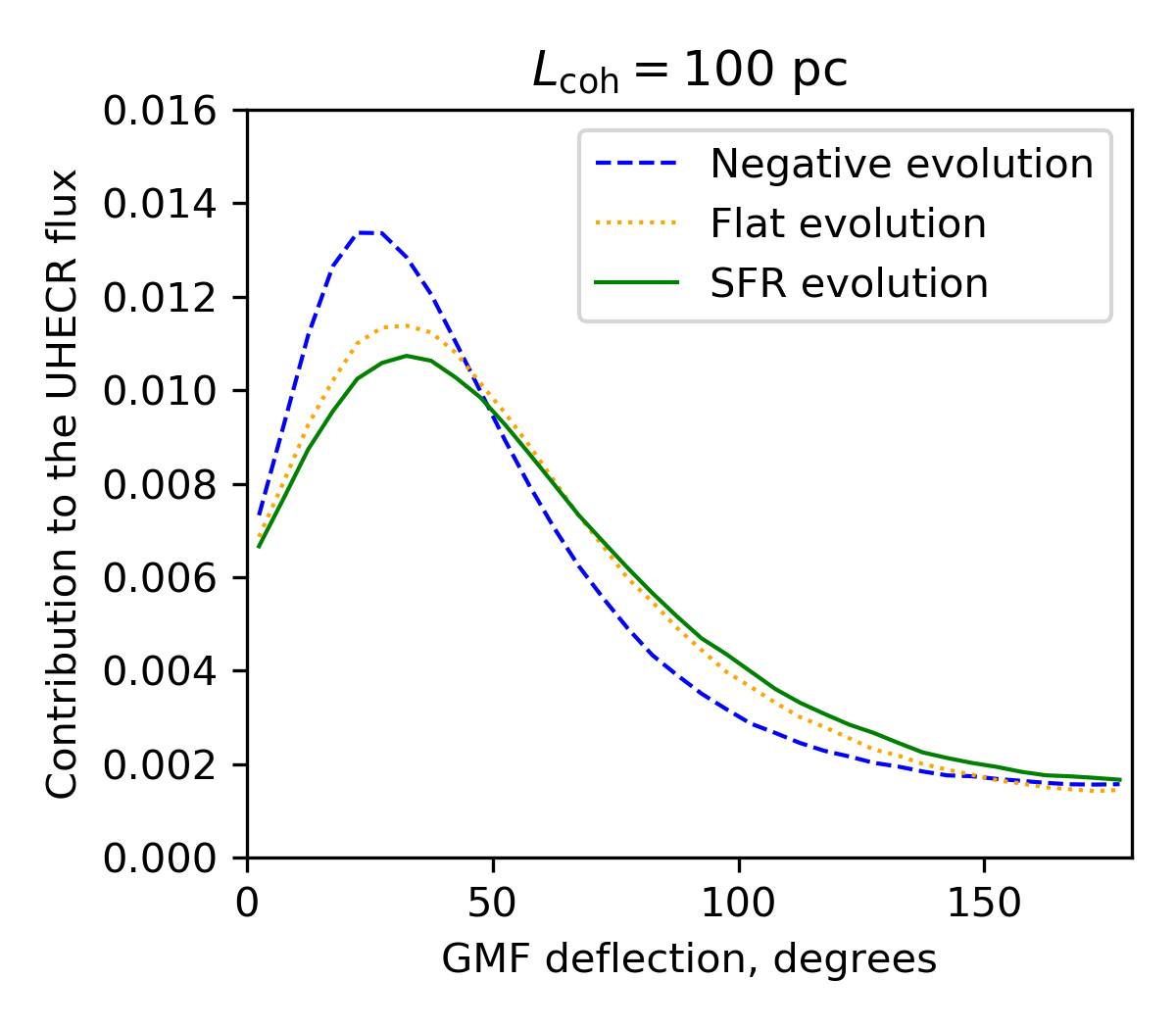}
	\includegraphics[width=0.4\textwidth]{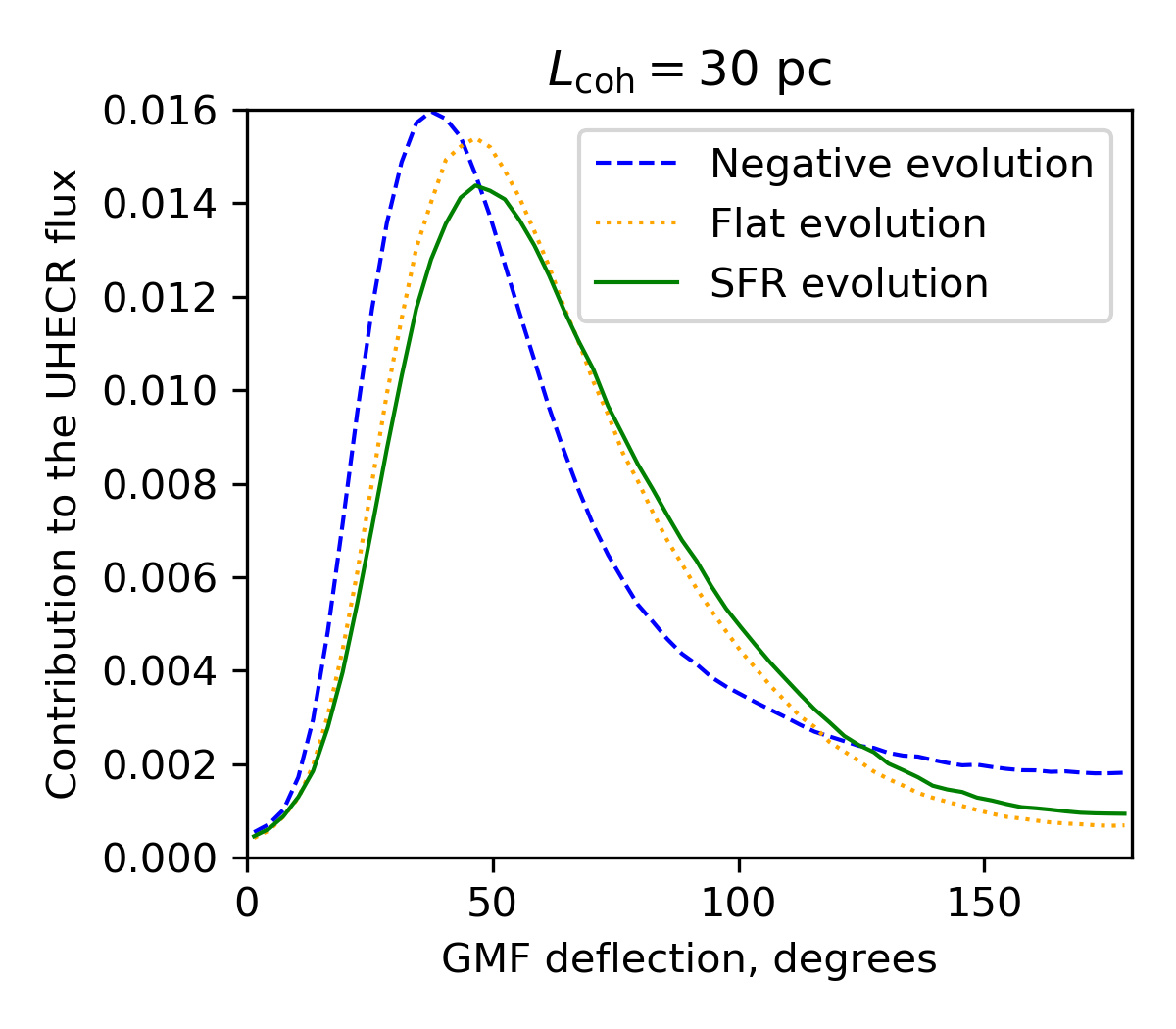}
	\caption{Distribution of deflections in the GMF for all UHECRs with $E_\text{CR}>10^{18.5}$~eV and for the three different source evolution scenarios: negative evolution (blue, dashed), flat evolution (orange, dotted) and sources following the SFR (green, solid). Left panel: the coherence length of the random component of the GMF $L_\text{coh} = 100$~pc. Right panel: $L_\text{coh} = 30$~pc. The results have been normalized so that the area under all curves is equal to one.}
	\label{plot:GMFdefl}
\end{figure*}

Besides the deflections in EGMFs the cosmic rays will also be deflected in the GMF. To estimate the GMF deflections we used the distribution of deflections as a function of the rigidity ($R$) as parametrized by \citet{Farrar:2017lhm}, taking into account deflections from different directions all over the sky. This distribution is parametrized in terms of the mean deflection of arrival directions ($\langle \Delta_{\text{GMF}} \rangle (R)$) and the RMS arrival direction spread ($\sigma(\Delta_{\text{GMF}}) (R)$)  for the JF12 GMF model \citep{Jansson:2012pc, Jansson:2012rt}, with $L_\text{coh} = 100$~pc and $L_\text{coh} = 30$~pc as correlation lengths for the random component of the GMF. The overall best-fit values for the parameters of these deflection distributions, as a function of $R$, are given there by
\begingroup
\small
\thinmuskip=\muexpr\thinmuskip*5/8\relax
\medmuskip=\muexpr\medmuskip*5/8\relax
$$
	\log_{10} \left( \frac{\langle \Delta_{\text{GMF}} \rangle}{\text{deg}} \right) = (-0.65 \pm 0.03) \log_{10} \left( \frac{R}{\text{V}} \right) + (13.63 \pm 0.59) \, ,
$$
$$
	\log_{10} \left( \frac{\sigma(\Delta_{\text{GMF}})}{\text{deg}} \right) = (-1.17 \pm 0.04) \log_{10} \left( \frac{R}{\text{V}} \right) + (23.22 \pm 0.71)
$$
\endgroup
for $L_\text{coh} = 100$~pc and
\begingroup
\small
\thinmuskip=\muexpr\thinmuskip*5/8\relax
\medmuskip=\muexpr\medmuskip*5/8\relax
$$
	\log_{10} \left( \frac{\langle \Delta_{\text{GMF}} \rangle}{\text{deg}} \right) = (-0.73 \pm 0.04) \log_{10} \left( \frac{R}{\text{V}} \right) + (15.23 \pm 0.70)  \, ,
$$
$$
	\log_{10} \left( \frac{\sigma(\Delta_{\text{GMF}})}{\text{deg}} \right) = (-1.03 \pm 0.03) \log_{10} \left( \frac{R}{\text{V}} \right) + (20.30 \pm 0.64)
$$
\endgroup
for $L_\text{coh} = 30$~pc. Each particle arriving at Earth in the 1D CRPropa simulations, which were also used to obtain the contribution to the UHECR flux as a function of the redshift (Fig.~\ref{plot:nuevo} lower right panel), is assigned a certain deflection randomly following the deflection distribution for its specific rigidity as parametrized by these equations. In this way a full distribution of GMF deflections for sources all over the sky is obtained based on the JF12 GMF model for each specific source-evolution scenario. The resulting distributions of GMF deflections are given in Fig.~\ref{plot:GMFdefl}.

\subsection{Monte Carlo simulation}
\label{sec:montecarlo}

Our Monte Carlo simulation, that is necessary to evaluate under which conditions we expect correlations between high energy neutrinos and UHECRs, is based on the following steps:

\textit{i)} We extract a number of sources $N_s$ at isotropically distributed random positions in the sky, with distances according to the source evolutions shown in the upper left panel of Fig.~\ref{plot:nuevo}.

\textit{ii)} We assign the probability to observe a neutrino from a certain source  proportional to $f_\nu(z)$ (see upper right panel of Fig.~\ref{plot:nuevo}), while we determine the probability to observe a cosmic ray based on its distribution  $f_{\text{CR}}(z)$ (see lower right panel of Fig.~\ref{plot:nuevo}).

\textit{iii)} We extract 36 neutrinos, corresponding to the number of observed through-going muons \citep{Haack:2017mfm}, which are the neutrino events characterized by a good directional resolution of roughly $1^\circ$ and an energy threshold of 200 TeV, to avoid most of the atmospheric background.\footnote{Concerning neutrinos, we are implicitly assuming that all the 36 through-going muons detected by IceCube are of extragalactic origin, neglecting the possible atmospheric background. This hypothesis maximizes the expected number of correlations between neutrinos and UHECRs. If the atmospheric background is present (or the Galactic plane region is excluded) the possibility to observe correlations gets worse.}. The positions of these neutrinos are random and they do not reflect the positions of through-going muons detected by IceCube. Then, we extract 135k cosmic rays, roughly the number of cosmic rays in the combined spectrum detected by Auger for $E_{\text{CR}}>10^{18.5}$~eV~\citep{Fenu:2017hlc}. 
Moreover, we are not using the true positions in which cosmic rays and high-energy neutrinos have been detected. Instead, both neutrinos and cosmic rays are chosen to arrive from randomly picked sources in the set of $N_s$ sources, with probabilities proportional to $f_\nu(z)$ and $f_{\text{CR}}(z)$. Therefore, cosmic rays and neutrinos have different positions in each simulations.

\textit{iv)} We include that neutrinos follow a straight path, while cosmic rays are deflected in EGMFs according to the function $\Delta_{\text{EGMF}}(z)$. In the lower left panel of Fig.~\ref{plot:nuevo} $\langle \Delta_{\text{EGMF}} \rangle (z)$ is illustrated. In the analysis the full distribution of $\Delta_{\text{EGMF}}(z)$ is used, which has a large spread (comparable with the median value) not represented in Fig.~\ref{plot:nuevo}. Additionally, the cosmic rays are deflected in the GMF following the distributions given in Fig.~\ref{plot:GMFdefl}. We implement here the distribution for $L_\text{coh} = 100$~pc as this choice gives the smallest deflections on average and is, therefore, expected to give the most correlations.

\textit{v)} We count the number of cosmic rays within a certain angular distance from the neutrino position. Performing a parameter scan we find that the ideal angular window is $5^\circ$ assuming the presence of EGMFs only, while this window is much larger (close to $20^\circ$-30$^\circ$) when also the GMF is included (see App.~\ref{app:angularwindow} for details about this parameter scan). The number of cosmic rays within the angular window represents the sum of signal and background ($s+b$). Then we count the average number of cosmic rays within the same angular window from random positions, which represents the background ($b$).

\textit{vi)} We sum the number of cosmic rays detected in all angular windows, and we compare this number with the total number of cosmic rays expected in the same angular windows in case of an isotropic distribution. Using the Poissonian likelihood
$$
\chi^2=2(b+s) \ln(1+s/b)-2s \, ,
$$
we consider that the correlations between neutrinos and UHECRs can be discovered if $\chi^2 > 25$ (corresponding to $5 \sigma$). 

\textit{vii)} We repeat the entire process $10^3$ times for a given set of number of sources and source evolutions. At the end of the process roughly $10^6$ sky maps are considered. For each number of sources and each source evolution, we compute how many maps among the $10^3$ maps show significant correlations (according to the definition given above), tagging the map as significant if the excess is larger than 5$\sigma$.
At the end of the process we evaluate for which local densities (for each source evolution) the median $5\sigma$ discovery potential is reached, meaning that at least 50\% of the maps show more than $5\sigma$ of excess. 

Note that the number of contributing sources $N_s$ will be critical for the results. This number translates into a local source density depending on the source evolution, see Appendix \ref{app:number} for details. For example, the same $N_s$ will lead to a very high local source density for negative source evolution, compared to an about $10^4$ times smaller value for the local source density for SFR evolution. Thus, while the presented results will apparently depend on local source density and source evolution, the implied dependence on $N_s$ is actually moderate. 

\begin{figure*}
\centering
\includegraphics[scale=0.3]{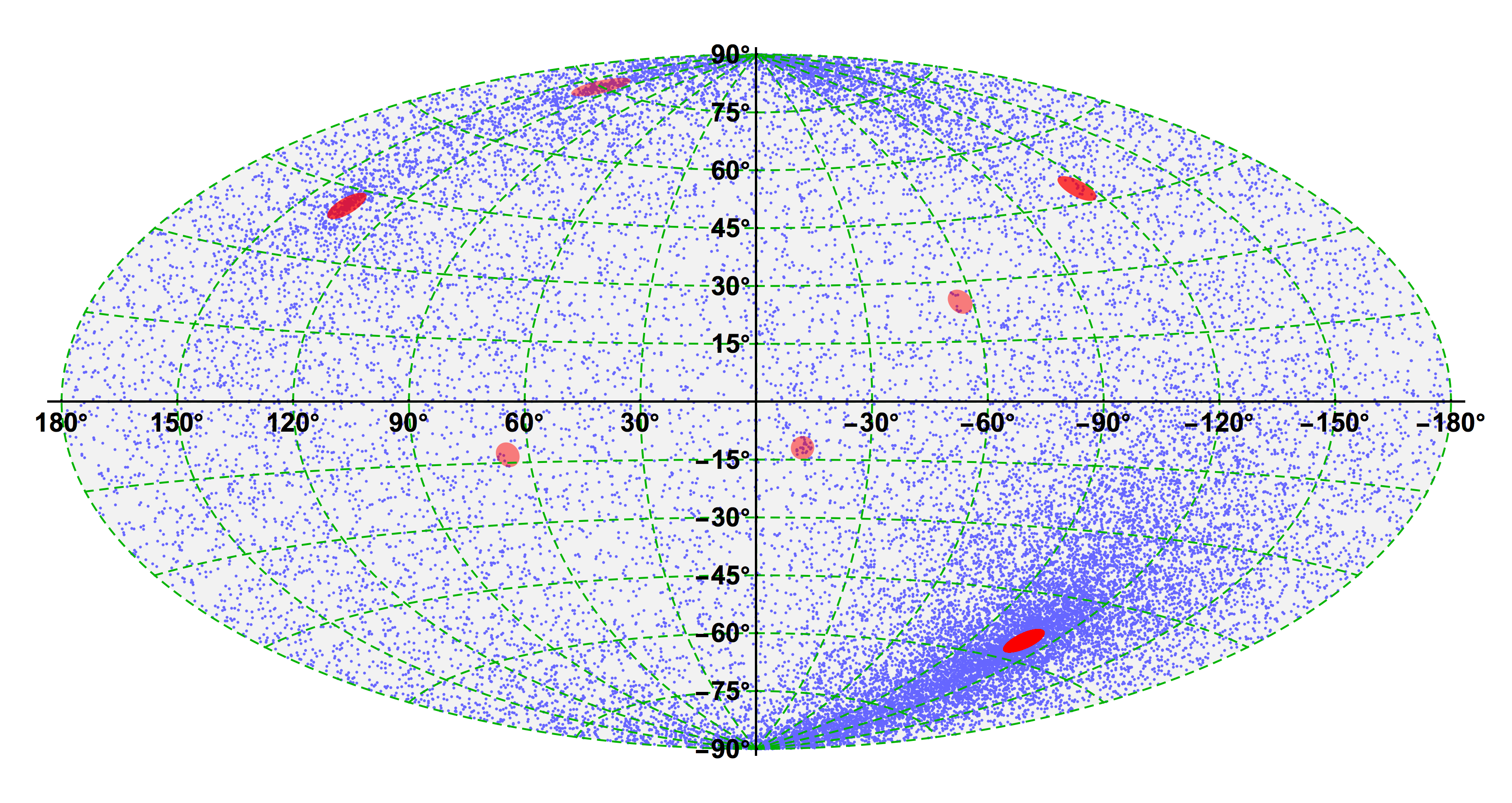}
\includegraphics[scale=0.3]{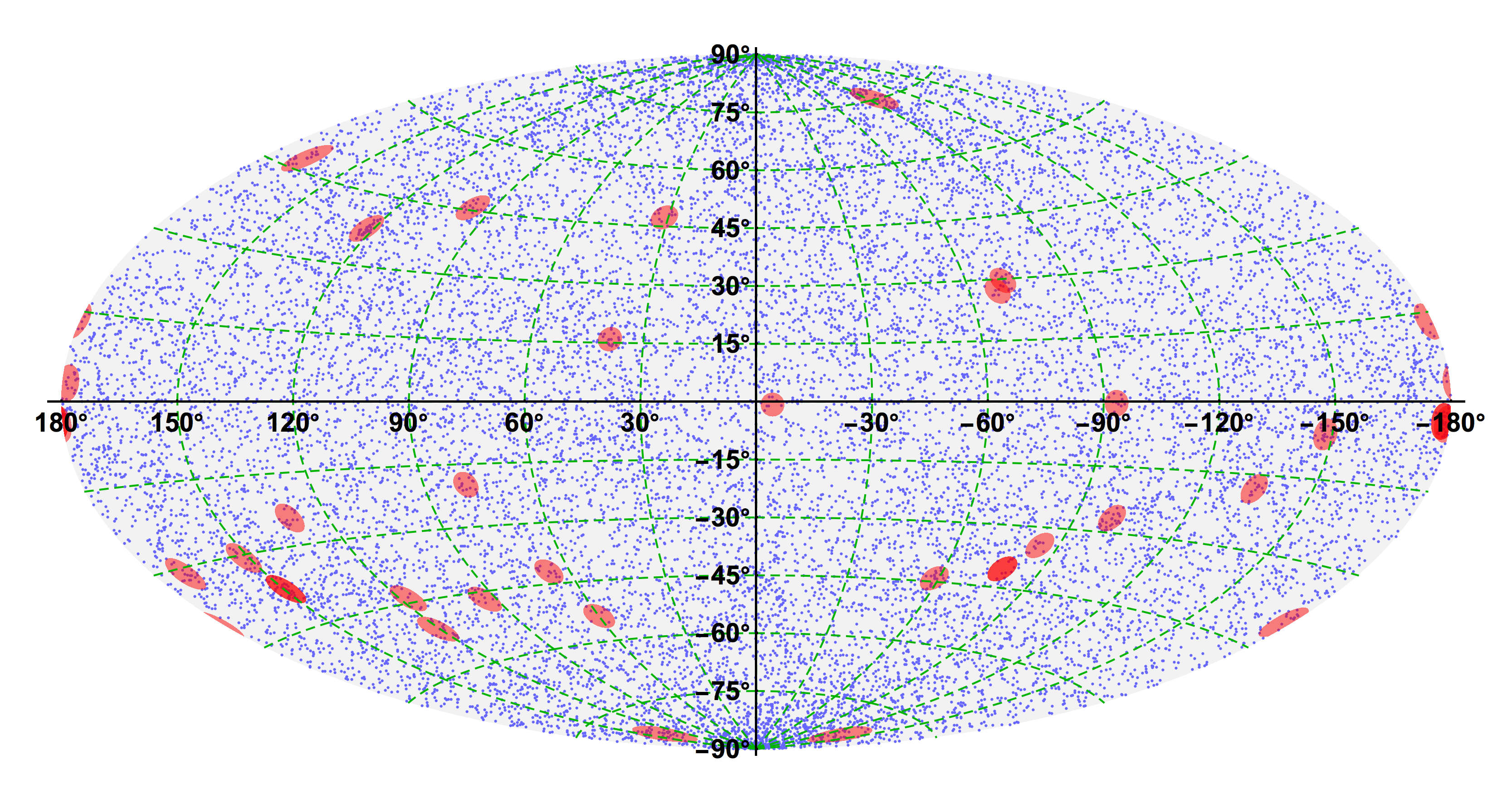}
\caption{Two different sky maps, in which neutrinos are represented with red disks (for an exaggerated uncertainty of $3^\circ$, much larger than the typical $1^\circ$ angular uncertainty for tracks) and cosmic rays are represented with blue points. In each sky map 36 neutrinos are shown (the number of neutrinos in the current IceCube through-going muon sample~\citep{Haack:2017mfm}), picked at random from $N_s$ sources, together with $10^5$ cosmic-ray arrival directions 
randomly selected from the same $N_s$ sources. We assume a negative source evolution, $E_{\text{CR}} \geq 10^{19}$~eV and $\rho_0=0.1$ and  $100$~Gpc$^{-3}$yr$^{-1}$ for the left and right panel respectively (corresponding to 8 and 8500 sources).
Note that the left sky map shows less than 36 disks because in this case multiple neutrinos arrive from the same sources. Furthermore, note that the directional resolution, the energy threshold and the number of cosmic rays are exaggerated here compared with our analysis to visualize the effect.
}
\label{fig:skymaps}
\end{figure*}

We illustrate our procedure in Fig.~\ref{fig:skymaps}, where we show two different sky maps for neutrinos (red disks) and cosmic rays (blue dots) for different local source densities and a negative source evolution; this means that the number of contributing sources $N_s$ scales directly with the local source density. For this specific case, the number of sources is equal to 8 and 8500 in the left and right sky maps, respectively. The neutrino error circles use an angular window of $3^\circ$ (for illustration purpose only) and several neutrinos may be detected in the same error circle (when the number of sources is very low).  Note that the energy threshold  is exaggerated to $E_{\text{CR}} \geq 10^{19} \mbox{ eV}$ (for illustration purpose only), while in the rest of the paper we use $E_{\text{CR}} \geq 10^{18.5} \mbox{ eV}$. Let us remark that these simulations only represent one possible realization for each number of sources.

Our procedure will establish in which cases an excess of cosmic rays within the red disk error circles can be established over the isotropic background, summed over all neutrino events.
While such an excess is obvious in the left panel of Fig.~\ref{fig:skymaps}, the other case requires a dedicated statistical analysis. It is also interesting to observe that smaller local source densities lead to a stronger clustering of both the cosmic rays and neutrinos (several neutrinos may fall into the same disk), which implies already that measuring the neutrino-UHECR connection and observing neutrino multiplets are correlated problems; see the next section.

\subsection{Neutrino multiplets}
\label{sec:numult}

The non-observation of neutrino multiplets strongly constrains the local source density under the assumption that the source class powers the diffuse neutrino flux \citep{Kowalski:2014zda,Ahlers:2014ioa,Murase:2016gly} -- an assumption which we also use. From the previous discussion it is qualitatively clear that a small number of contributing sources $N_s$ implies both a significant UHECR anisotropy and a high probability for neutrino multiplets. It can, therefore, be expected that the non-observation of neutrino multiplets limits the parameter space to find neutrino-UHECR correlations. 

We use the same procedure as the one outlined for neutrino-UHECR correlations to determine the probability to observe neutrino multiplets, as a function of the number of sources and the source evolution. It is expected that for a small number of sources this probability is close to unity, such as for FSRQs -- which are very bright but rare $\gamma$-ray sources. On the other hand, when sources are abundant and less luminous (such as starburst galaxies), the probability to observe neutrino multiplets is close to zero for the current exposure. 

For consistency, we determine the parameter space in which the probability to observe a neutrino multiplet is larger than 90\%. That is obtained by repeating the Monte Carlo simulation for $10^3$ different local densities and the three different cosmic evolutions, producing $10^3$ sky maps for each case. Then we count in how many simulations neutrino multiplets are present -- varying the source luminosity, the local density and the source evolution. It is important to note that the multiplets are sensitive to the total number of contributing sources $N_s$ as well, which then translates into the local density depending on the source evolution.
 
Note that our counting statistics procedure differs from the methods in the literature such as \citet{Murase:2016gly,Ackermann:2019ows}. Our results agree with the IceCube analyses for SFR evolution \citep{Aartsen:2018ywr,Aartsen:2018fpd}, while they appear to be more conservative than the results presented in \citet{Murase:2016gly,Ackermann:2019ows}. However, note that given the different assumptions and methods we re-compute the multiplet constraint ourselves. Then we compare the region excluded by the absence of multiplets with the region in which correlations between UHECR and high energy neutrinos are expected, which was obtained using the same assumptions and methods. Furthermore, note that all of these analyses (including ours) assume that the entire neutrino flux is powered by one single source class, which means that sources in the excluded regions can potentially still power a fraction of the observed neutrino flux.

\section{Results}
\label{sec:results}

\begin{figure*}
\centering
\includegraphics[width=0.32\textwidth]{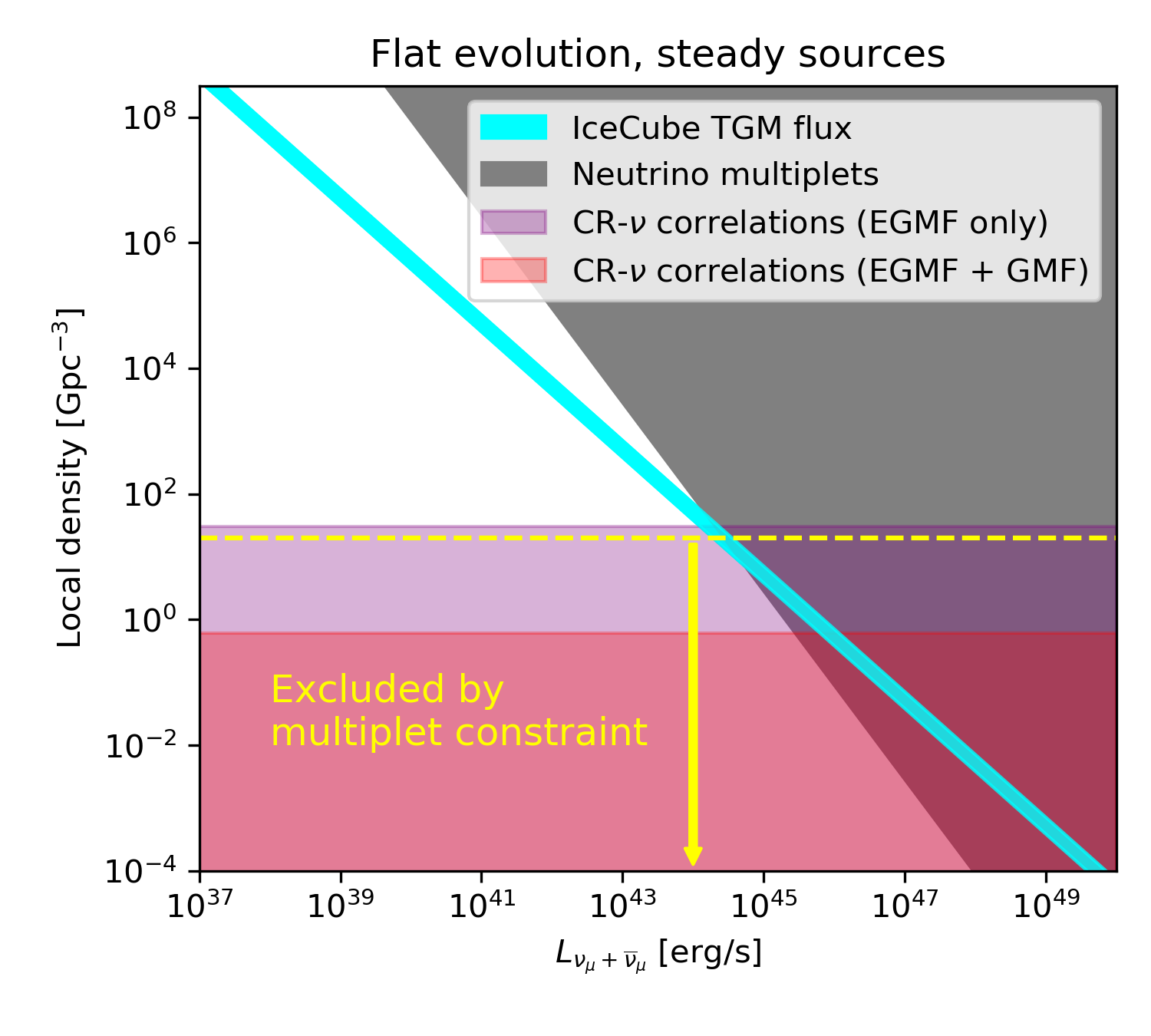} 
\includegraphics[width=0.32\textwidth]{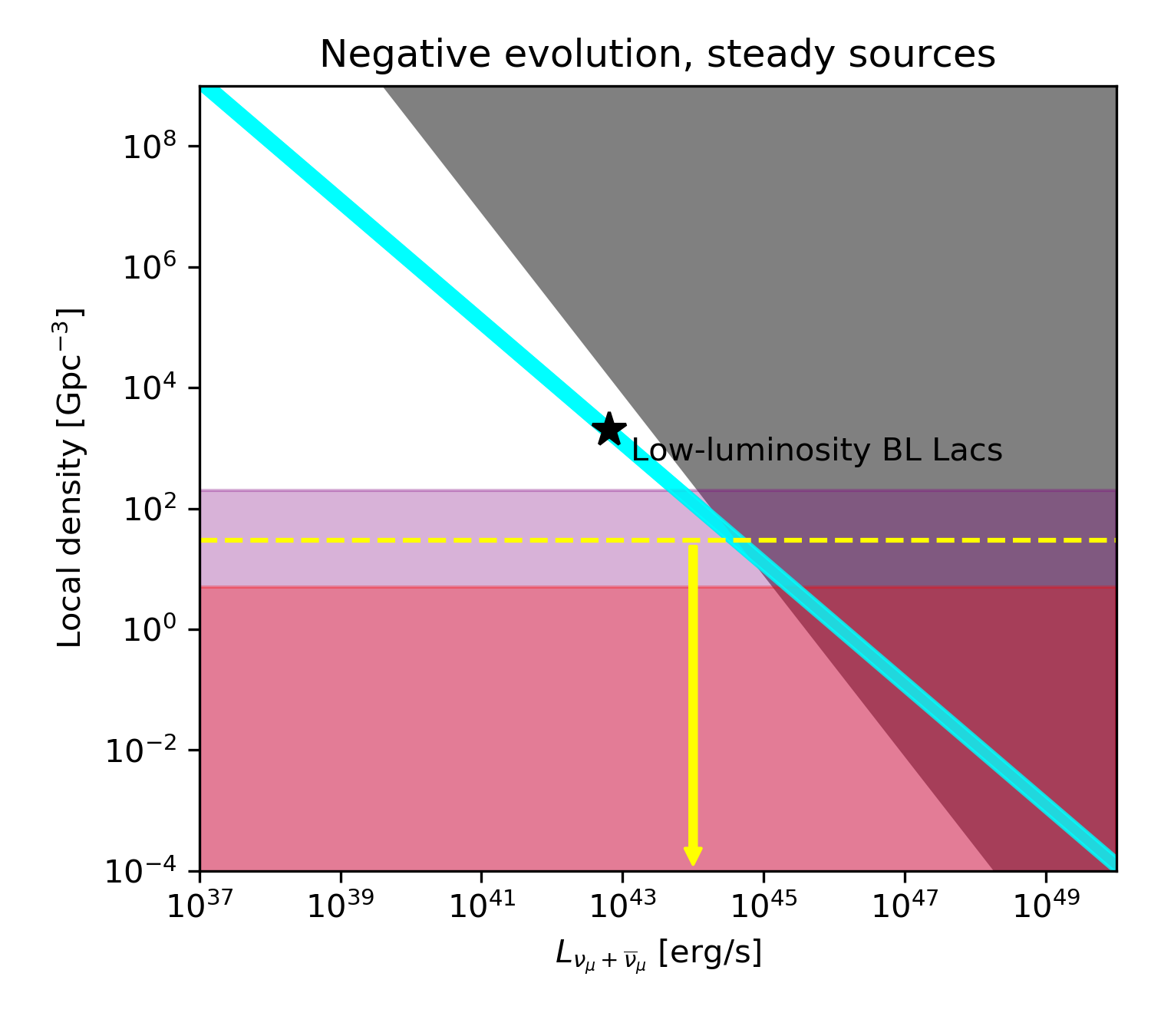} 
\includegraphics[width=0.32\textwidth]{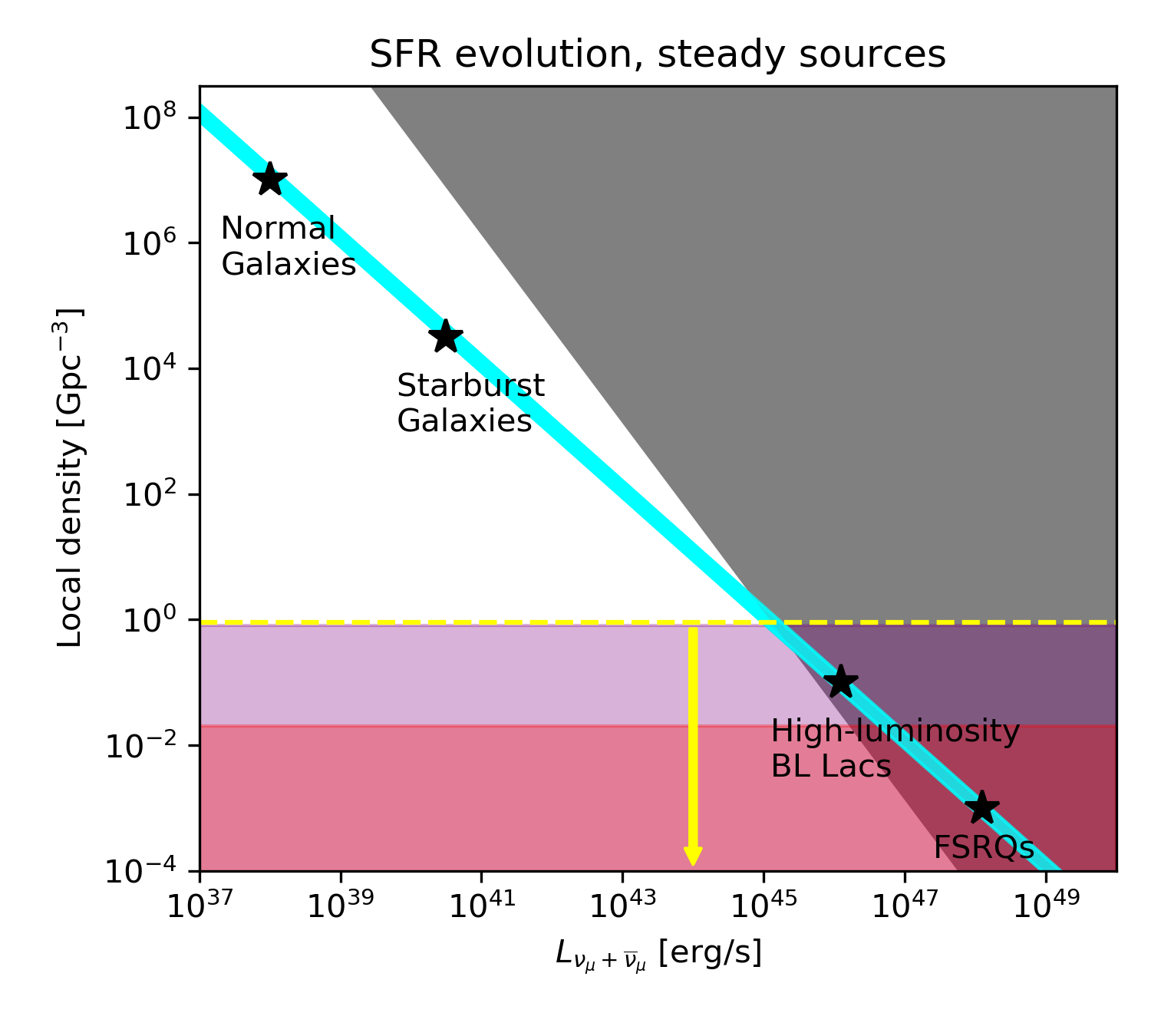}  \\
\includegraphics[width=0.32\textwidth]{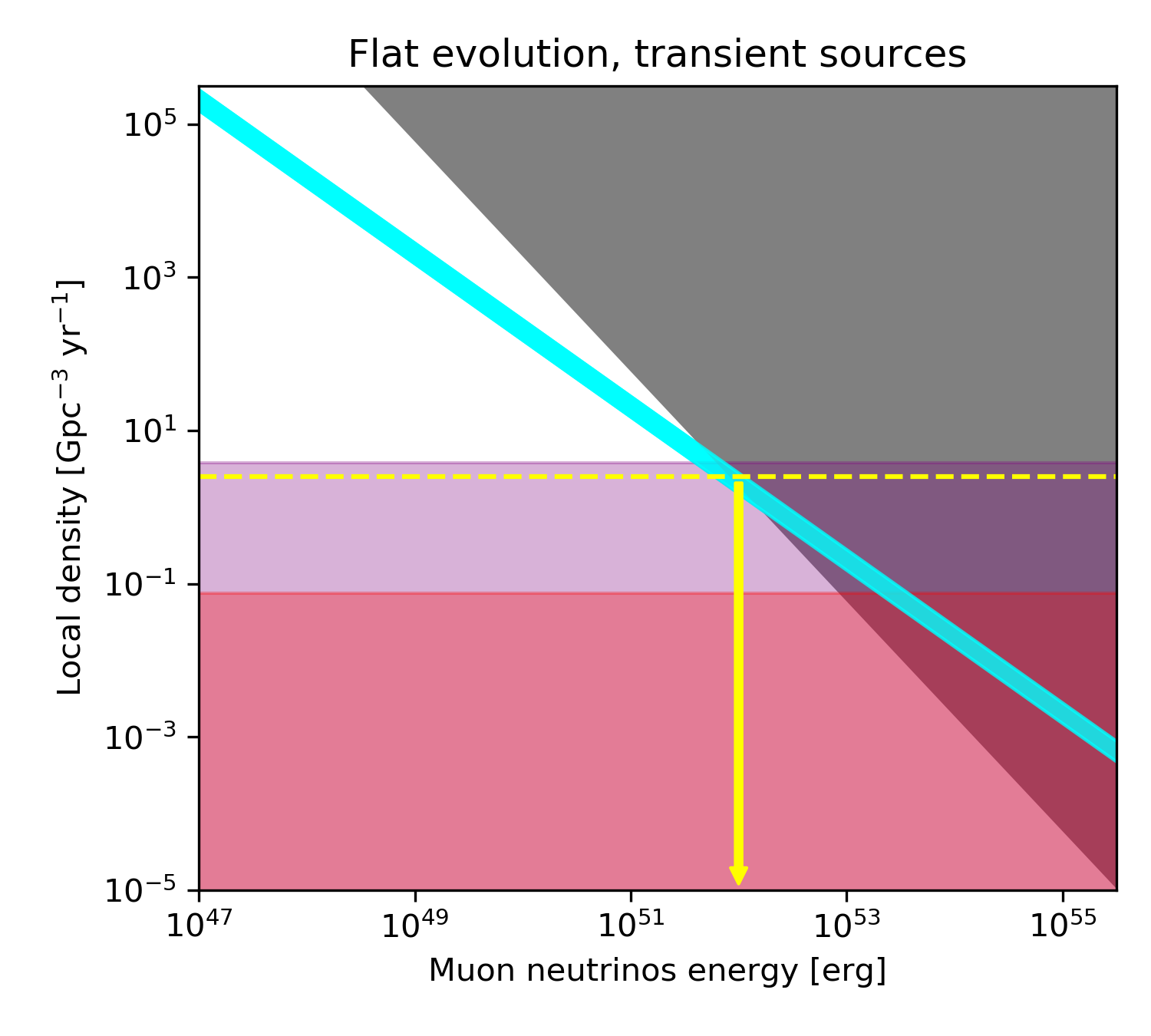} 
\includegraphics[width=0.32\textwidth]{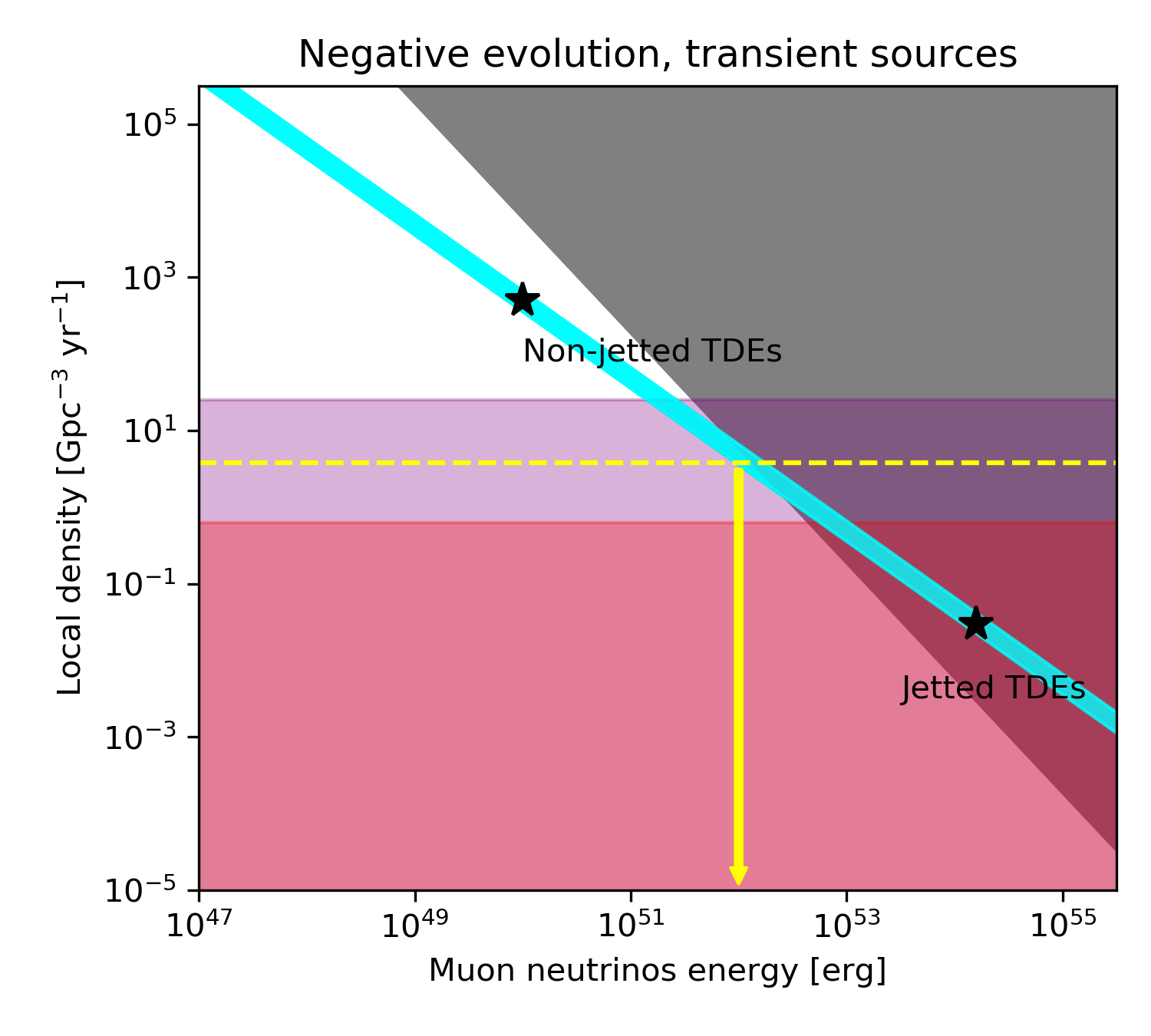} 
\includegraphics[width=0.32\textwidth]{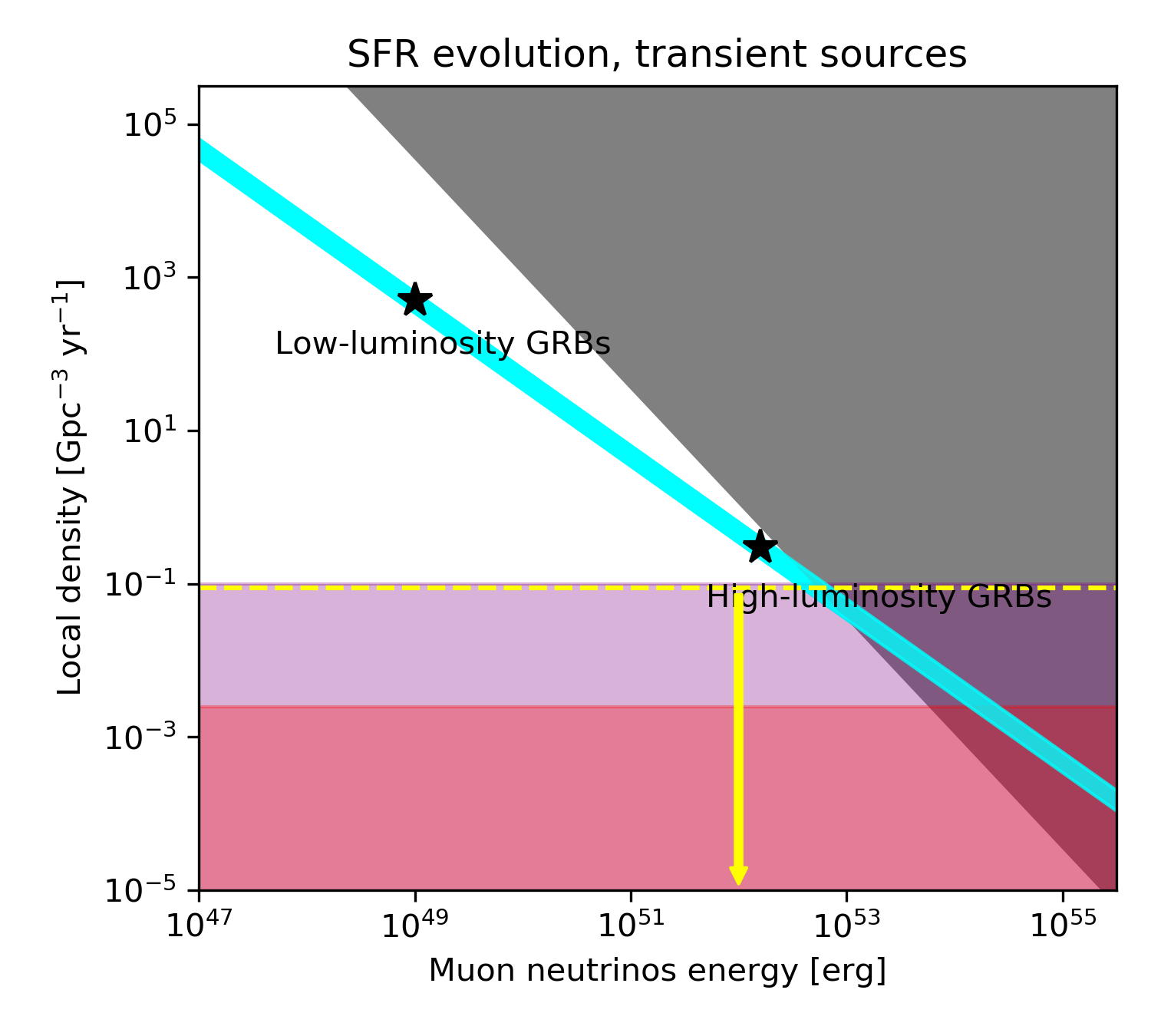}  
\caption{Neutrino multiplet constraints and the neutrino-UHECR constraints represented in the same figure.
The cyan regions represent the local luminosity density required to power the through-going muon flux measured by IceCube after eight years of data taking \citep{Haack:2017mfm}. The gray regions represent the parameter space excluded by the non-observation of neutrino multiplets for the same exposure. Consequently, local densities below the yellow lines, where these two regions intersect, are excluded. On the other hand, discovering the neutrino-UHECR correlation requires low source densities below the purple (red) shaded regions excluding (including) the GMF in addition to the EGMF; the median $5\sigma$ discovery potential is shown.  
Some typical sources, according to their evolution, are shown using stars. The different panels refer to different source evolutions and steady or transient sources, as indicated.}
\label{fig:finres}
\end{figure*}

We show our main result in Fig.~\ref{fig:finres} in the form in which typically the neutrino multiplet limits are shown, for different source evolutions and steady or transient sources in different panels. The difference between steady and transient sources is represented by an extra factor, for transient sources, of $(1+z)^{-1}$ in Eq.~(\ref{eq:probnu}) and the same extra factor in the conversion between local density and number of contributing sources (see Appendix~\ref{app:number}). 

In the cyan region,  the product between the average source luminosity $L_\nu$ (computed between $10^{4}$ and $10^7$~GeV) and the local density $\rho_0$ matches the through-going muon flux detected by IceCube; this can be interpreted as an effective local luminosity density. This is obtained using a power-law flux of $E_\nu^{-2.2}$ for each source (as suggested by through-going muons \citep{Haack:2017mfm}), assuming that all the sources are standard candles with the same luminosity. The diffuse neutrino flux is then given by the following expression:
$$
\frac{\text{d}\phi_\nu}{\text{d}E_\nu} = \int_0^{z_{\text{max}}} \frac{D_H }{4 \pi \ h(z)} \frac{\mathcal{L}(E_\nu(1+z))}{E_\nu^2 (1+z)^2} \rho(z) \text{d}z \, ,
$$
where $\mathcal{L}(E_\nu)$ is a power-law function reproducing the source luminosity when integrated between $10^4$ and $10^7$~GeV, while $D_H$ and $h(z)$ are defined in Appendix~\ref{app:number}. The function $\rho(z)$ denotes the source evolution, as represented in Fig.~\ref{plot:nuevo} upper left panel.
We then require that the diffuse flux is equal to the measured through-going muon flux, resulting in a fixed value for the product between the luminosity of the standard-candle sources and the local density of the sources (depending on their evolution).

In the same figure, the gray regions denote the regions excluded by the absence of neutrino multiplets in 8 years of collected through-going muon data by IceCube. Particularly, in this region the probability to observe neutrino multiplets is larger than 90\%.

Finally, the regions above the purple areas have source densities that are too large for the detection of neutrino-UHECR correlations with the current sensitivities of IceCube and Auger. Concerning the purple and red regions, we always assume that there are 36 neutrinos (i.e.\ that the luminosity is the one required to power the through-going muon flux). 
Consequently, the allowed parameter space in which a neutrino-UHECR correlation can be detected is marked as purple and red regions.
The purple region is the one obtained considering only the presence of the EGMF, while the red region is the one obtained considering also the presence of the GMF, which increases the average deflection.

From Fig.~\ref{fig:finres}, it can be read off that the possibility to observe correlations between neutrinos and UHECRs is restricted to a small region of the parameter space of $L_\nu$ and $\rho_0$, when only the EGMF is accounted for. In fact, if such a correlation were found, the combination with the neutrino multiplet constraint would give a precise measurement of the local source density and the source luminosity. The largest parameter space remains for a negative source evolution (middle panels), since in that case both observed particle species are produced in the nearby Universe; compare right panels in Fig.~\ref{plot:nuevo}. A detection of the neutrino-UHECR connection would mean that the source density (for steady sources, negative evolution) has to be around 10--1000 $\mathrm{Gpc^{-3}}$, and the luminosity around $10^{44}$ to $10^{45} \, \mathrm{erg \, s^{-1}}$. Using the total number of sources as a free parameter (instead of the local density) we find that the number of sources required to observe correlations $N_s \lesssim 10^4-10^5$. This number depends weakly on the source evolution, while the local density is strongly affected by the source evolution. Indeed, fixing the total number of sources, the local density is roughly 200 times smaller for SFR-like evolution compared to the negative evolution (see Appendix~\ref{app:number}). 

However, when also the GMF is considered (choosing the most optimistic case, i.e.,\ the left panel of Fig.~\ref{plot:GMFdefl} that produces on average a smaller deflection), the possibility to observe correlations becomes weak, since the required densities are already in tension with the absence of multiplets in neutrino data. Let us remark that we have also optimized the analysis, changing the angular window for this case, namely going from $5^\circ$ (EGMF only) to $30^\circ$. Indeed, in order to observe correlations the total number of sources must not be too high, while at the same time the absence of multiplets in neutrino data suggests the opposite. 

\section{Discussion}
\label{sec:discussion}

The results presented in the previous section are obtained under certain assumptions, that are in most of the cases optimistic for finding correlations between UHECRs and neutrinos. Even under these assumptions the results, considering deflections in both the EGMF and the GMF, are already quite pessimistic. In addition, there are some elements that we are neglecting and that can diminish the expected number of neutrino-UHECR correlations even further.

First of all, we assume that the detection probability of both cosmic rays and neutrinos is equal across the entire sky. This is an ideal assumption, since the sky coverage of realistic experiments is not taken into account. Concerning the IceCube neutrino telescope, for example, track-like events (the ones with a good angular resolution) come mostly from the northern hemisphere, while Auger has the highest sensitivity in the southern hemisphere. However, the inclusion of the sky coverages and detector sensitivities is beyond the purpose of this work, which aims to illustrate the possibility to observe correlations between neutrinos and cosmic rays under optimistic assumptions.

Second, we have included the GMF deflections from \citet{Farrar:2017lhm} using the most optimistic assumption (the smallest average deflection), represented in the left panel of Fig.~\ref{plot:GMFdefl}. The uncertainties on the GMF are still very large and if the true GMF is close to the one represented in the right panel the situation will get worse. Another point to note concerning the GMF is that the deflections shown in Fig.~\ref{plot:GMFdefl} represent the best-fit values averaged over the full sky. In \citet{Farrar:2017lhm} is shown that, in general, different deflections are expected when looking at different directions through the Galaxy. For example, according to the parametrizations given in \citet{Farrar:2017lhm}, the mean separation from the source direction for $R = 10^{18.5}$~V for the $L_{\mbox{coh}} = 100$ pc case is $45.7\degree$ in the northern hemisphere, $46.8\degree$ in the southern hemisphere, $28.8\degree$ in the Galactic plane and $40.3\degree$ in the full sky. Additionally, there are strong variations in the expected deflections depending on the exact position in the sky, making the separation in northern hemisphere, southern hemisphere and Galactic plane rather arbitrary. As our analysis evaluates the discovery potential for neutrino-UHECR correlations for all events in the full sky combined, we chose to rather use the deflection distribution for the entire sky instead of separating in different sky regions.  

Third, for the EGMF the weakest model of the six models described in \citet{Hackstein:2017pex} has been chosen. The authors of that paper have tried to span the full range of allowed EGMF models, from very weak to very strong EGMFs. The weakest model they found is, therefore, a suitable choice for the most optimistic scenario for finding UHECR and neutrino correlations. The uncertainties on the EGMF are even larger than on the GMF and a different choice of EGMF model could give significantly larger deflections and, therefore, less expected neutrino-UHECR correlations. The expected deflections in the case of strong EGMFs are discussed in \citet{AlvesBatista:2017vob}, where they show that average deflections of up to $90\degree$ can be reached even for protons with energies close to $10^{20}$~eV. Such strong fields would significantly reduce the expected neutrino-UHECR correlations. 

Fourth, Eq.~(\ref{eq:probnu}) is valid if the neutrino flux is power-law distributed. However, a power-law flux it is not an universal spectrum expected in all situations. In case neutrinos are produced by the interactions between accelerated protons and background photons, the resulting neutrino energy spectrum has a typical bump shape, that differs significantly from the power-law behavior.  In this case it is important to verify that the peak of the energy spectrum is inside the region in which the detector is efficient. 
If this condition did not apply, sources that produce cosmic rays might produce neutrinos that are not detectable by the present experiment, which would reduce the number of effective common sources in our analysis. 
This is equivalent to the assumption that only a fraction of neutrinos and cosmic rays come from the same source class; in both cases the expected number of neutrino-UHECR correlations goes down. This is natural to understand, since the result would correspond to the one presented in this work with a reduced exposure, depending on the fraction of common sources. 

\citet{Heinze:2019jou} shows that the UHECR fit to the spectrum and mass composition can change significantly when choosing different models for hadronic interactions (EPOS~LHC \citep{eposlhc}, Sibyll~2.3 \citep{Riehn:2017mfm} or QGSJet~II-04 \citep{qgsjetii}) in UHECR air showers. While a different choice of one of these models results in changes in the expected source evolution and spectral indices, the maximum rigidity and the composition at the sources are better constrained. We treat the uncertainty in the source evolution here by showing the results for three different scenarios, with corresponding best-fit parameters for the spectral index, maximum rigidity and composition. The expected deflections in the EGMF mainly depend on the UHECR composition at the sources, which is predicted rather robustly according to \citet{Heinze:2019jou}. The deflections in the GMF model depend on the UHECR composition at our Galaxy, which differs depending on the choice of hadronic-interaction models. The best-fit results we used all assumed EPOS~LHC as interaction model. On average, EPOS~LHC predicts a lighter composition (and therefore less deflections in the GMF) than Sibyll~2.3, but a heavier composition than QGSJet~II-04. However, QGSJet~II-04 does not produce a consistent relation between cosmic-ray mass and $X_\text{max}$ variables \citep{Aab:2014kda, Bellido:2017cgf} and could, therefore, be considered as disfavored. We have, however, also tested an extreme scenario where we consider a pure-proton composition. Even in that case no correlations between UHECRs and neutrinos are expected when deflections in the GMF are included (see Appendix~\ref{app:proton} for these results).

Besides cosmic ray-neutrino correlations and neutrino-neutrino correlations, limits on the local source density can also be obtained from cosmic ray-cosmic ray correlations. This is what the Pierre Auger Collaboration has done in \citet{Abreu:2013kif}. The limits obtained there show that $\rho_0 \gtrsim (0.06 - 7) \times 10^{5}$~Gpc$^{-3}$, even stronger than the limits from the neutrino multiplet constraints, making it even less likely that neutrino-UHECR correlations will be found. However, the analysis presented there assumes a maximum deflection of $30\degree$, while in our case often larger deflections than that are obtained (see Fig.~\ref{plot:nuevo} bottom left panel and Fig.~\ref{plot:GMFdefl} left panel). In addition, a different minimal energy threshold for cosmic rays is considered in \citet{Abreu:2013kif} compared with our results ($E_{\text{CR}} \geq 60 \mbox{ EeV}$ versus $E_{\text{CR}} \geq 10^{18.5} \mbox{ eV}$, respectively). It should also be noted that the bounds in \citet{Abreu:2013kif} were derived from the lack of significant clustering in arrival directions of the highest energy events detected at the Pierre Auger Observatory between 1 January 2004 and 31 December 2011. By now, the number of detected events at these energies has increased significantly and hints of anisotropies have been found~\citep{Aab:2018chp, Caccianiga:2019hlc}, which could affect the bounds on the density of UHECR sources.

Throughout this work we have used a minimal cosmic-ray energy threshold of $E_{\text{CR}} \geq 10^{18.5} \mbox{ eV}$, which is roughly the energy of the \lq\lq ankle\rq\rq . This choice of minimal energy maximizes the number of detected UHECRs while still making sure that most UHECRs can be expected to have an extragalactic origin and that the best-fit UHECR spectrum and composition results (Table~\ref{tab:CRbestfitparam}) are valid. In \citet{Aartsen:2015dml,Schumacher:2019qdx}, however, 52\,EeV and 57\,EeV are used as minimal UHECR energy thresholds for events detected by Auger and TA, respectively. Such a higher energy threshold will reduce the expected deflections of UHECRs in magnetic fields. However, the maximum source distance from which UHECRs can arrive at Earth is also reduced significantly (limiting the neutrino-UHECR even further), and the number of detected UHECRs becomes much less. To investigate the effect of a higher minimal UHECR energy threshold we have redone the entire analysis described in this paper for $E_{\text{CR}} \geq 50 \mbox{ EeV}$, in which case the chance of detecting neutrino-UHECR correlations only becomes smaller, see Appendix~\ref{sec:CR50} for these results.  

\section{Summary and conclusions}
\label{sec:conclusion}

We have scrutinized the question if the neutrino-UHECR connection is, in principle, detectable based on correlations of arrival directions. To investigate the most favorable scenario, we have assumed that all of the neutrinos in the through-going muon sample, which have excellent directional resolution, and all cosmic rays above $10^{18.5} \, \mathrm{eV}$ originate from the same source class. We have taken into account the different horizons of neutrinos and cosmic rays, the impact of the cosmological evolution of the sources, and deflections of cosmic rays in Galactic and extragalactic magnetic fields.

We have demonstrated that the problem of observing the neutrino-UHECR connection is intimately connected with the non-observation of neutrino multiplets. While the non-observation of neutrino multiplets implies that the number of contributing sources  has to be high enough (and the corresponding luminosity low enough) not to detect several neutrinos from the same source, the neutrino-UHECR connection requires a relatively small number of sources to produce the anisotropy related to the neutrino arrival directions. Consequently, the non-observation of neutrino multiplets limits the possibility to observe the neutrino-UHECR connection -- if it exists.

We have found that the best scenario for finding the neutrino-UHECR connection are sources with a negative source evolution. That is easy to understand: Since the UHECR horizon is limited by photo-hadronic interactions (such as photo-disintegration) with the cosmic background light, there is the most statistical overlap with the neutrinos (which can travel through the whole Universe) if the source evolution follows a similar trend. Consequently, a negative source evolution, such as it may be expected for (jetted or non-jetted) TDEs or low-luminosity BL Lacs, offers the best perspective. Even in that case, 
the allowed window on the source density (10--1000 $\mathrm{Gpc^{-3}}$) is relatively small if IceCube does not find neutrino multiplets, when only the extragalactic magnetic field is considered. Adding the contribution of the Galactic magnetic field, the required density to observe correlation has to be smaller than 10 $\rm Gpc^{-3}$ (negative evolution) -- in contradiction with the non-observation of neutrino multiplets. 

If, on the other hand, a significant neutrino-UHECR correlation is discovered, it will be an indirect precise measurement of the source density and a strong indication for a negative source evolution, and could thus lead to the identification of the source class -- and consequently to the discovery of the origin of cosmic rays. If, on the other hand, not even the proposed upgrade IceCube-Gen2 finds neutrino multiplets within a few years of operation, the neutrino-UHECR connection will not be detected in the near future.

We conclude that the perspectives for detecting the common origin of neutrinos and UHECRs are challenging. For example, for SFR evolution, the parameter space is already strongly constrained by the non-observation of neutrino multiplets. Even in the best scenario, i.e.\ negative source evolution, the potential  to observe correlations is limited to the  possibility to observe the first neutrino multiplet in the very near future. If IceCube does not observe any multiplet in the neutrino data in the next few years, searching for connection between neutrinos and UHECRs will become meaningless. 

\section*{Acknowledgements}

We would like to thank Andrew Taylor and Foteini Oikonomou for useful discussions and Julia Tjus and Xavier Rodrigues for commenting on the draft. This project has received funding from the European Research Council (ERC) under the European Union's Horizon 2020 research and innovation programme (Grant No. 646623) and the Initiative and Networking Fund of the Helmholtz Association.

\bibliographystyle{mnras}
\bibliography{bibliography}

\appendix
\section{Source evolution}
\label{sec:sevo}
In this section we analyze the contribution to the flux of neutrinos and cosmic rays as a function of redshift, assuming different cosmological evolutions of the sources, namely a negative, a flat and a SFR evolution.
We define these three different source evolutions per comoving volume $\rho(z)$, as follows:
\begin{itemize}
\item sources are characterized by having a negative source evolution, i.e.\ the amount of nearby sources is larger than the amount of distant sources. For this case we assume $\rho(z) \propto (1+z)^{-3}$, based on \citet{Sun:2015bda};
\item A flat evolution. The sources are uniformly distributed with redshift.
\item The source distribution follows the SFR. In this case the evolution is characterized by 
$$
\rho(z) \propto 
	\begin{cases}
	  (1+z)^{3.4} & \text{for } z < 1 \\
	  (1+z)^{-0.3} & \text{for }  1 \leq z \leq 4 \\
      (1+z)^{-3.5} & \text{for } z > 4
	\end{cases},
$$
according to \citet{Yuksel:2008cu}.  
\end{itemize}
The three source distributions $\rho(z)$ are represented in the upper left panel of Fig.~\ref{plot:nuevo} as a function of redshift. 

In order to compute the contribution to the neutrino flux, given a certain source distribution, we need to introduce some elements of cosmology.  First of all we define the comoving distance as:
$$
D_c(z)= D_H \times d(z) \, ,
$$
where $D_H= c/H_0$, with $H_0=67.3 \rm \ \frac{km/s}{Mpc}$ and $D_H= 4.46 \mbox{ Gpc}$ \citet{Aghanim:2018eyx}. The function $d(z)$ is defined as:
$$
d(z)=\int_0^z \frac{\text{d}z^{'}}{h(z^{'})} \, ,
$$
with 
$$
h(z) = \sqrt{\Omega_\lambda+ \Omega_m (1+z)^3}
$$
and $\Omega_\lambda=0.685$, $\Omega_m=0.315$, following \citet{Aghanim:2018eyx}.

Therefore, the comoving volume is given by:
$$
V_c= \frac{4}{3} \pi D_c^3 
$$
and the derivative in $z$ is equal to:
$$
\frac{\text{d}V_c}{\text{d}z} = 4 \pi D_H^3 \frac{d^2(z)}{h(z)} \, .
$$
In addition, we define the luminosity distance, that takes into account the redshift energy loss, as $D_\ell(z)=D_c(z) (1+z)$.

Given the source density per comoving volume $\rho(z)$, the contribution to the neutrino luminosity is 
given by:
$$
f_\nu(z)  = \tilde{f_\nu}^{-1} \rho(z) \frac{\text{d}V_c}{\text{d}z} \frac{1}{D_\ell^2(z)} \, ,
$$
with $\tilde{f}_\nu = \int_0^{z_{\text{max}}} \text{d}z \ \rho(z) \frac{\text{d}V_c}{\text{d}z} \frac{1}{D_\ell^2(z)}$, where we choose $z_{\text{max}}=5$ (the exact value chosen for $z_{\text{max}}$ is irrelevant for the final results as long as $z_{\text{max}} \gtrsim 3$). For transient sources the previous equations have an extra factor $(1+z)^{-1}$, as discussed in \cite{Kistler:2009mv}.
The functions $f_\nu(z)$, for the three different cases, are shown in the upper right panel of Fig.~\ref{plot:nuevo}.

\section{Number of sources versus local density}
\label{app:number}
The local density $\rho_0$ and the total number of sources $N_s$ are connected to each other and the constant of proportionality depends on the source evolution. The relation between these two quantities can be expressed as follows (see e. g. \citet{Baerwald:2014zga} for an extended discussion):
$$
\rho_0 = \frac{N_s}{4 \pi D_H^3 h_z} \, ,
$$
where
$$
h_z= \frac{1}{4\pi D_H^3} \int_0^{z_{\text{max}}} \frac{\rho(z)}{\rho(z=0)} \frac{\text{d}V_c}{\text{d}z} \tau(z) \text{d}z \, , 
$$
here we assume $z_{\text{max}}=5$, $D_H$ is the Hubble distance and $\rho(z)$ the source evolution as defined in Sec.~\ref{sec:sevo}. For steady sources $\tau(z)=1$  while $\tau(z)=1/(1+z)$  for transient sources.
The value of $h_z$ changes according to the source evolution and the topology of the source. In the case of steady sources, we obtain $h_z=0.1$ for negative evolution, $h_z=2.2$ for flat evolution and $h_z=18.0$ for SFR evolution. 

\section{Scan on the angular window}
\label{app:angularwindow}
\begin{figure}
\centering
\includegraphics[scale=0.5]{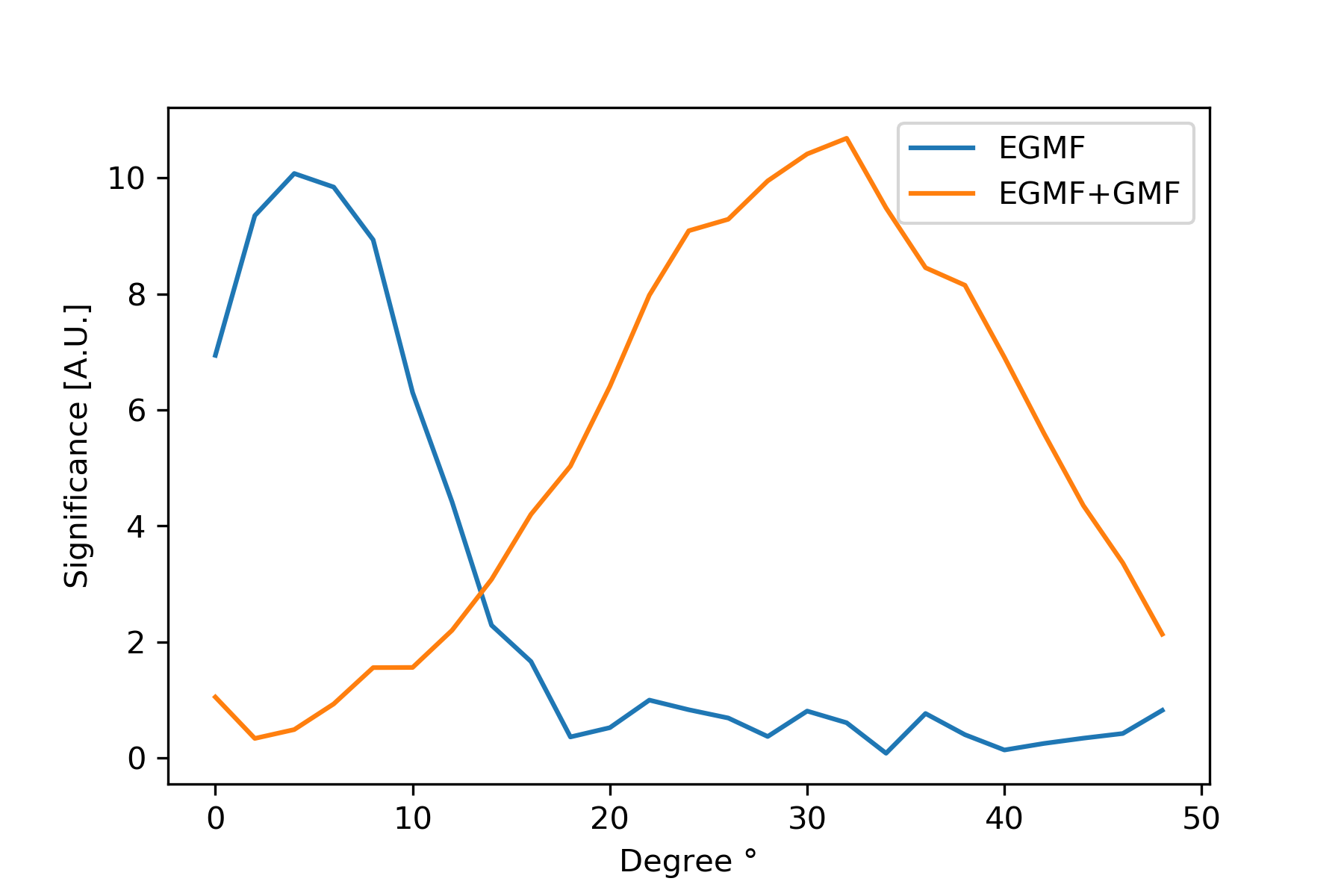}
\caption{Search for the optimal angular window to observe connections between UHECRs and high energy neutrinos, assuming deflections in the extragalactic magnetic field only (blue line) and adding the Galactic magnetic field (orange line).}
\label{fig:angasc}
\end{figure}

To determine what the optimal angular window is to search for correlations between UHECRs and high energy neutrinos, we perform a scan using different values for the angular window. We repeat the scan for two scenarios, i.e.\ extragalactic magnetic field only and extragalactic + Galactic magnetic fields. For this purpose we simulate $10^3$ sky maps, using SFR and a local density of 1 Gpc$^{-3}$, because in this scenario correlations are guaranteed (see upper right panel of Fig.~\ref{fig:finres}). See Fig.~\ref{fig:angasc} for the results of this scan.

\section{Pure-proton composition}
\label{app:proton}
\begin{figure*}
\centering
\includegraphics[width=0.32\textwidth]{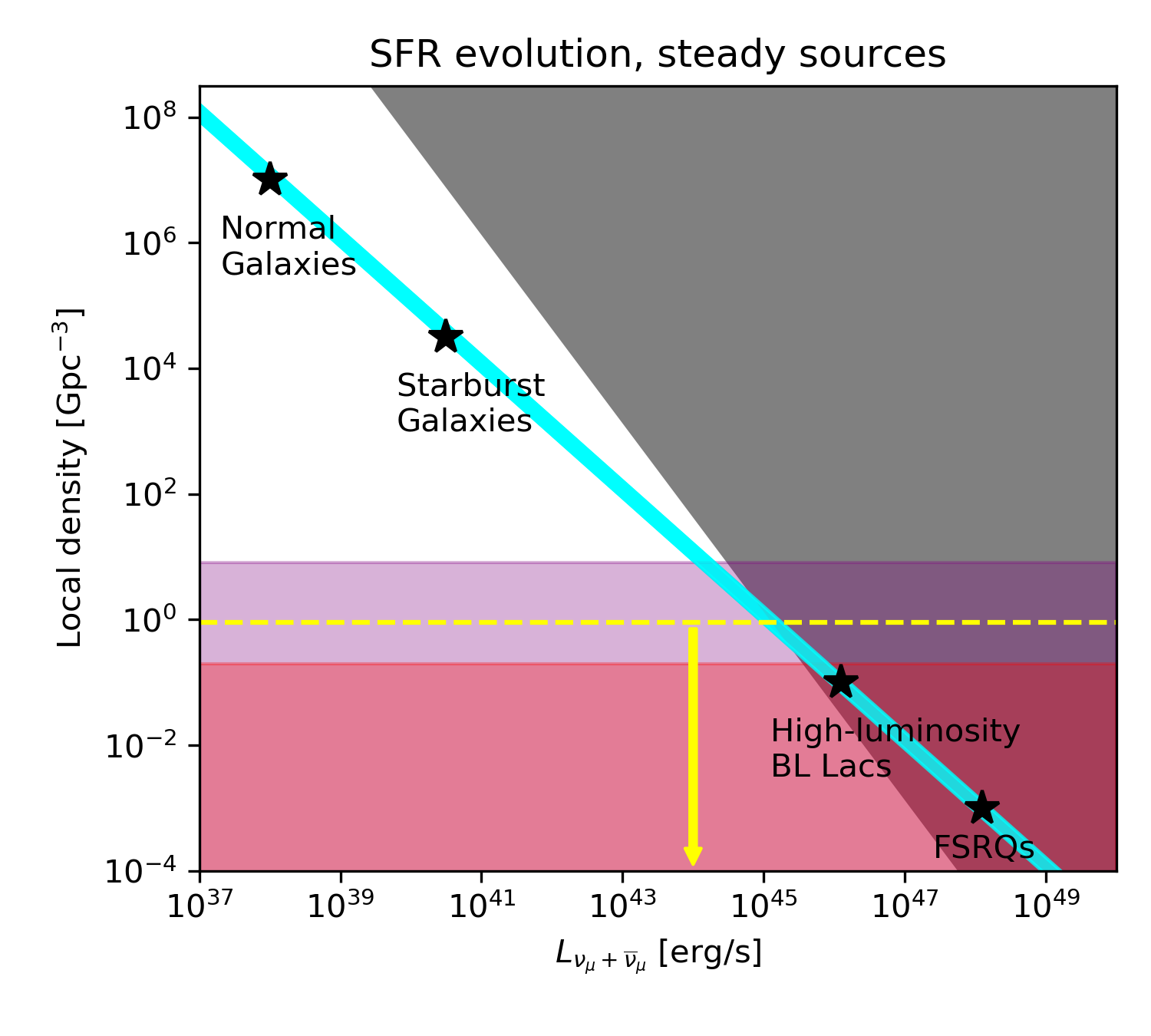}
\includegraphics[width=0.32\textwidth]{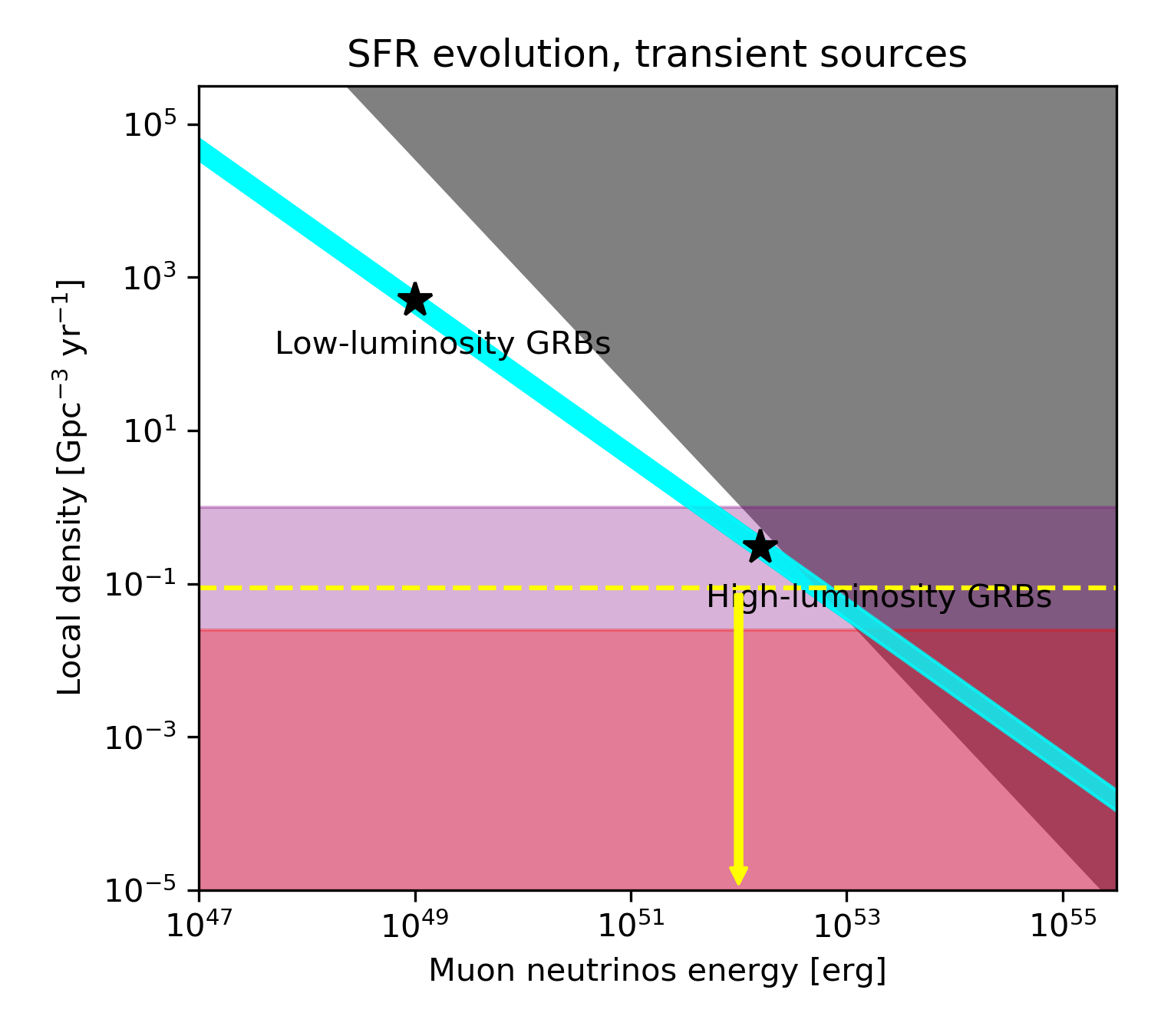}  
\caption{As in Fig.~\ref{fig:finres}, for a scenario with protons only and a minimal cosmic-ray energy threshold of $10^{18.5}$ eV. A SFR source evolution is implemented.}
\label{fig:finproton}
\end{figure*}

In this section we evaluate the effect of a different cosmic-ray composition, using purely protons and the same energy threshold used in the main text, i.e.\ protons above $10^{18.5}$ eV. Although this scenario has already been excluded by the Auger measurements, it is instructive as it is the most optimistic case in terms of UHECR-neutrino connections, since protons are deflected less than heavier nuclei. We did this for sources following the SFR with best-fit values for $\gamma = 2.42$ and $R_{\text{max}} = 10^{21.9}$~V obtained from \citet{Heinze:2015hhp}. The angular window for finding correlations between high energy neutrinos and UHECRs has been optimized for this specific scenario. For the EGMF only, the optimal angular window is 3$^\circ$, while for EGMF + GMF the optimal angular window is $15^\circ$. The results are represented in Fig.\ref{fig:finproton}. However, even in this extreme case, once the Galactic magnetic field is taken into account, the possibility to find correlations between UHECRs and high energy neutrinos in disfavored.

\section{Cosmic rays above 50~EeV}
\label{sec:CR50}
\begin{figure*}
\centering
\includegraphics[width=0.32\textwidth]{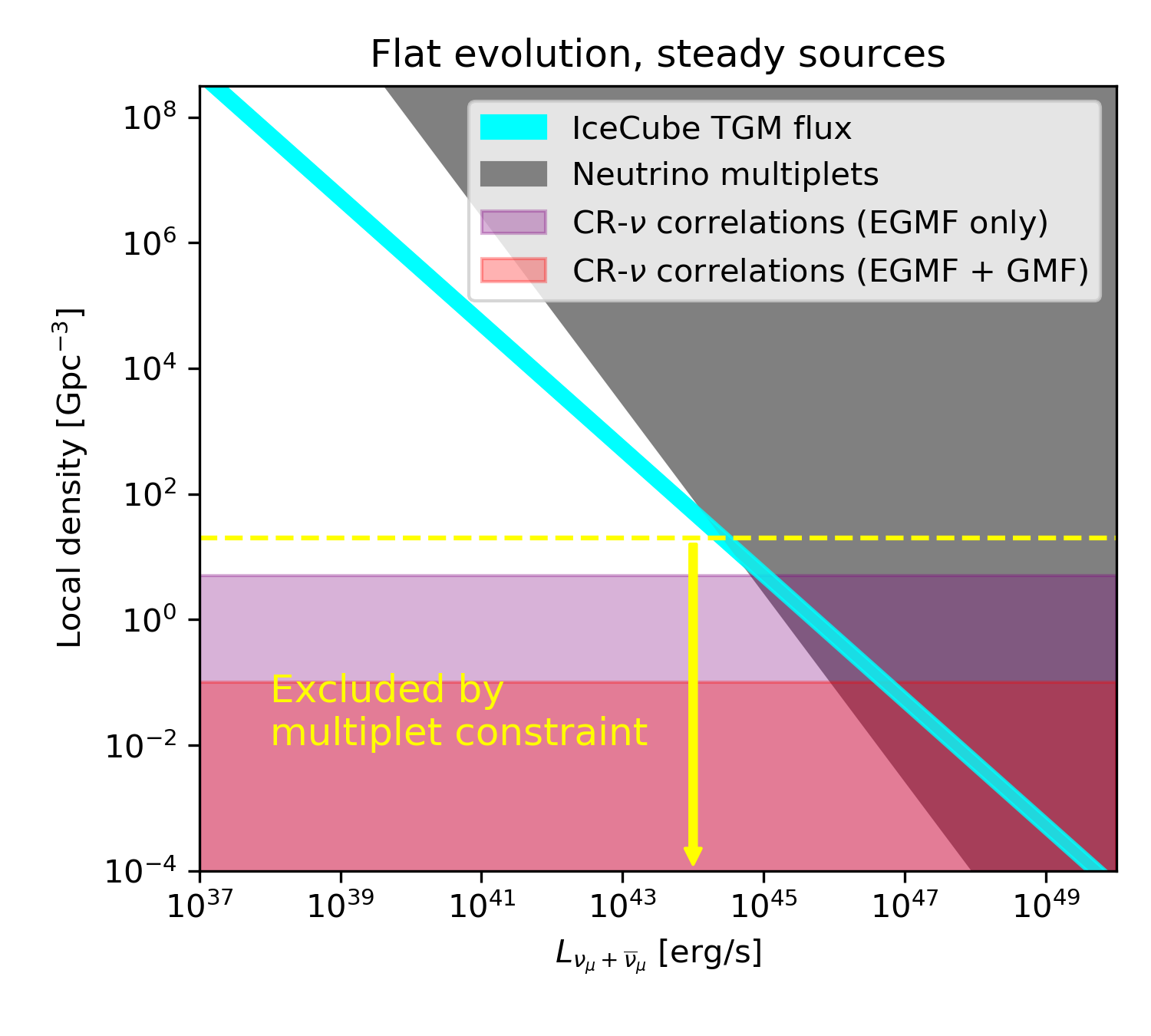} 
\includegraphics[width=0.32\textwidth]{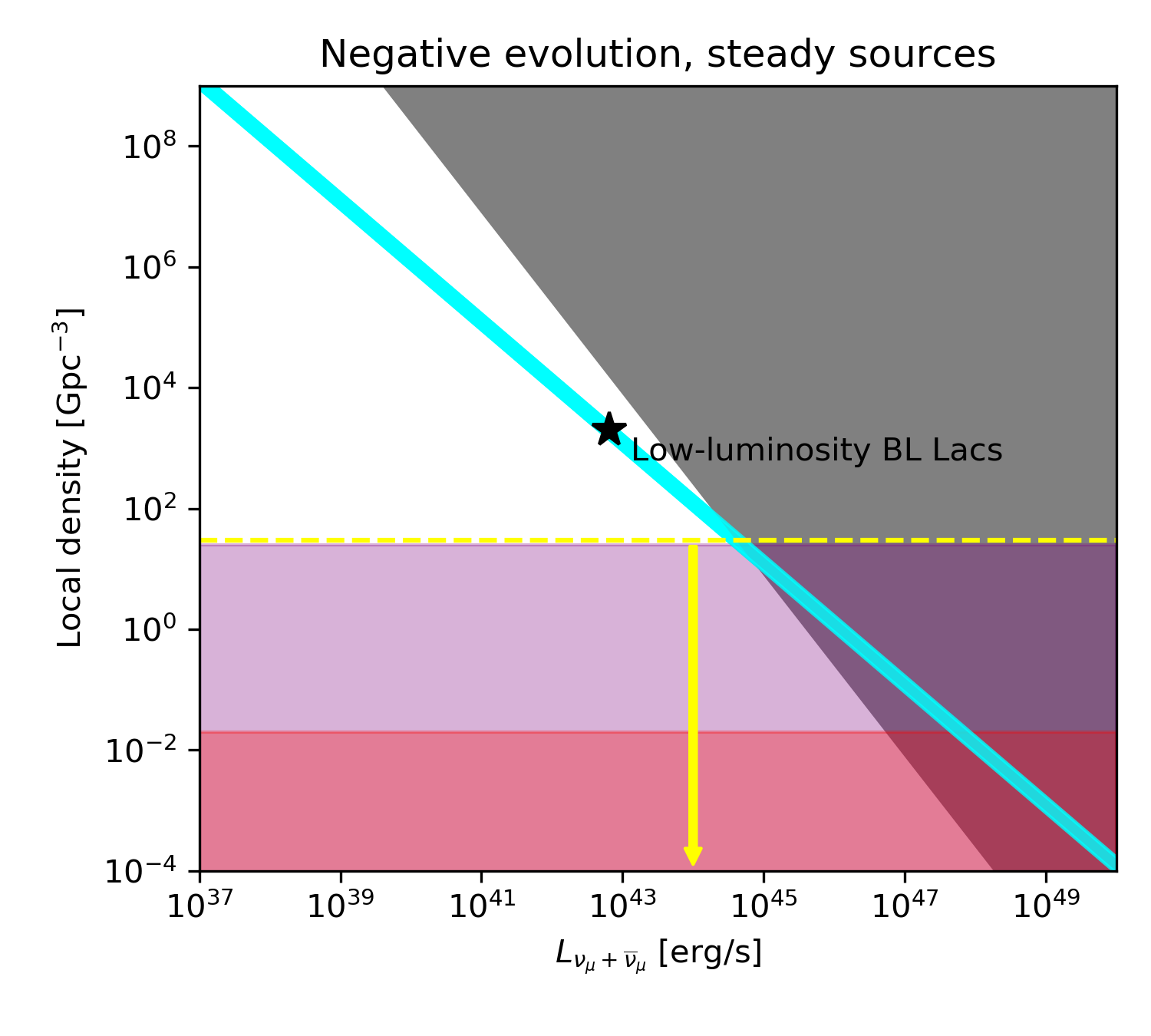} 
\includegraphics[width=0.32\textwidth]{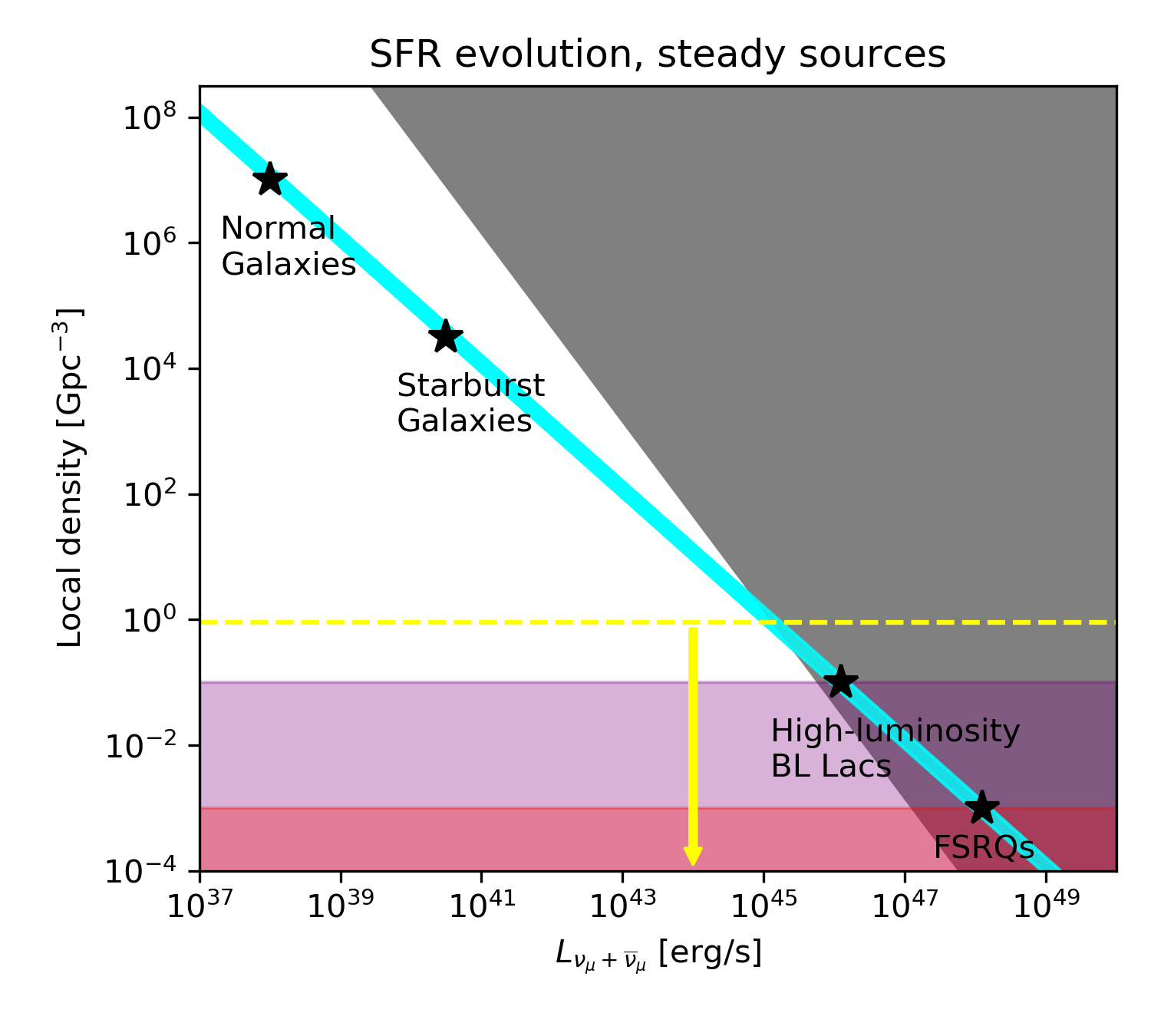}  \\
\includegraphics[width=0.32\textwidth]{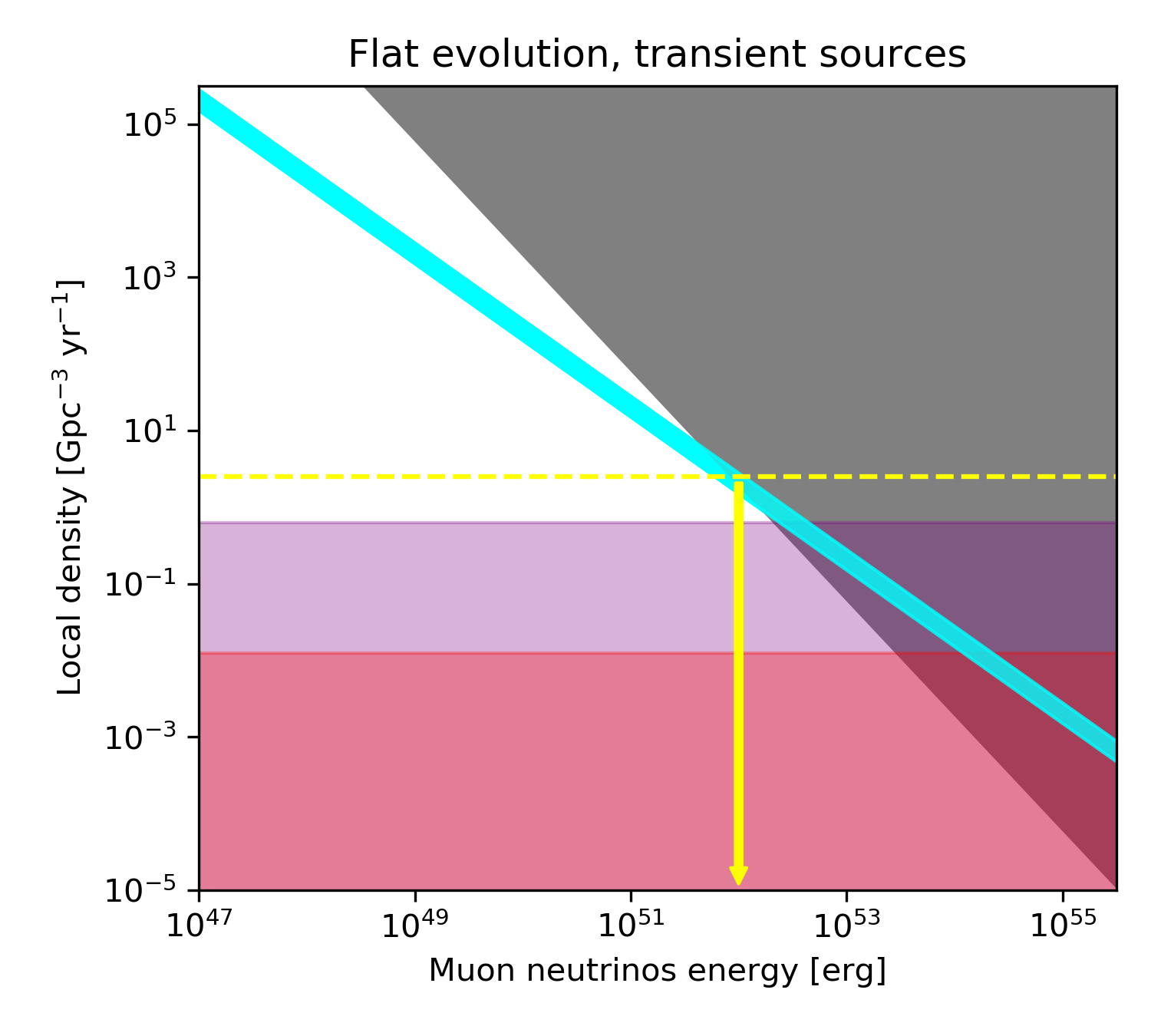} 
\includegraphics[width=0.32\textwidth]{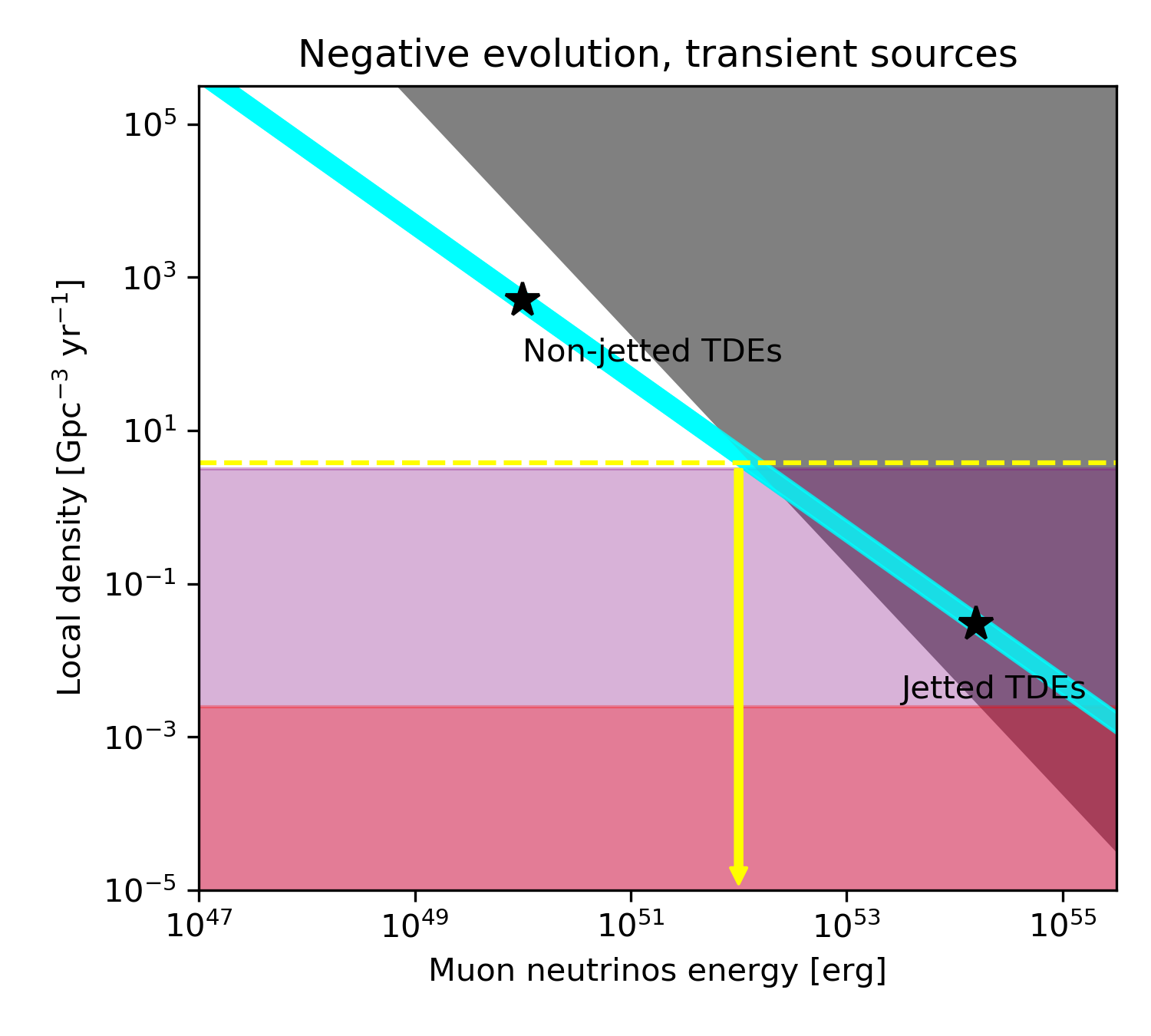} 
\includegraphics[width=0.32\textwidth]{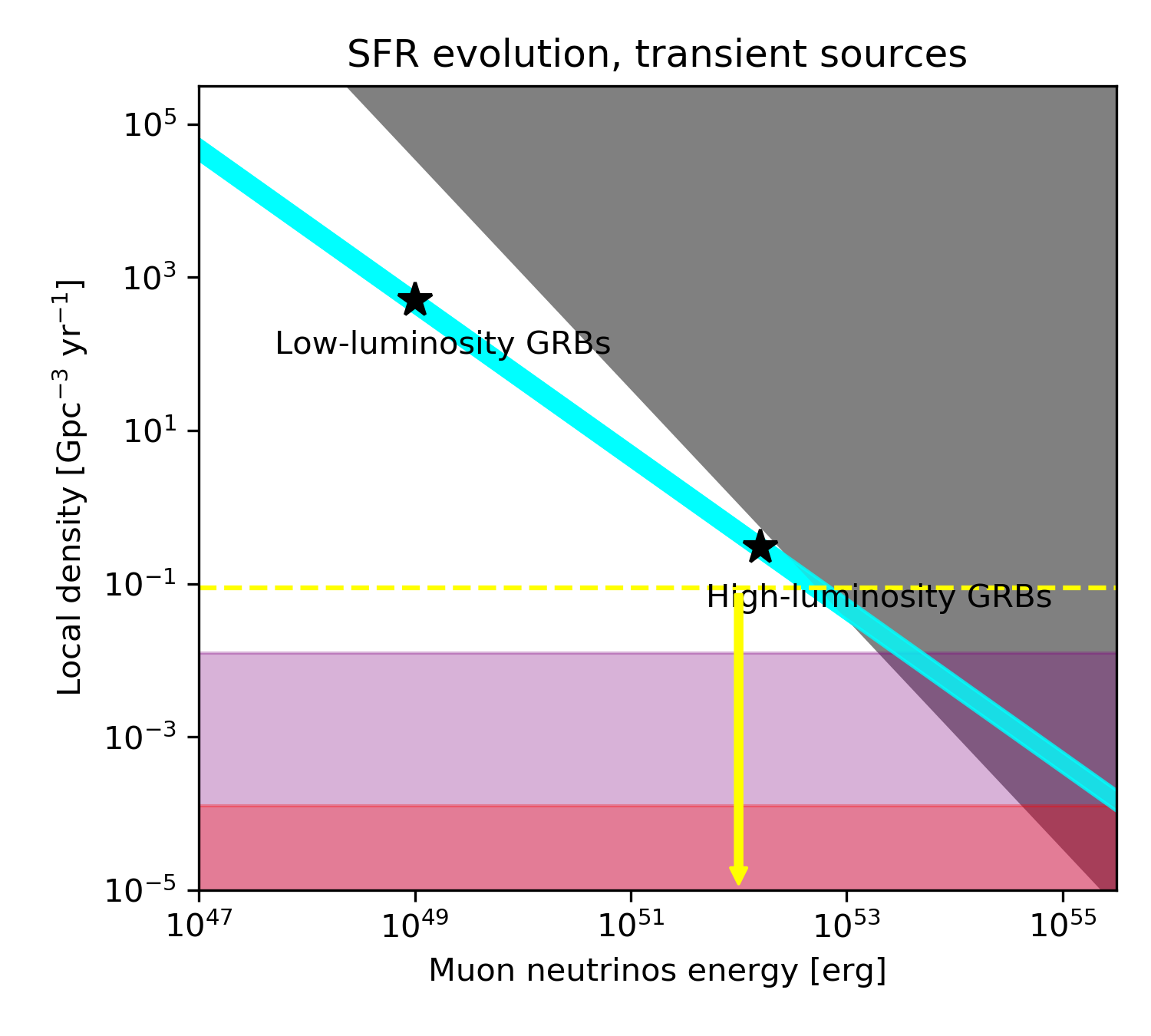}  
\caption{As in Fig.~\ref{fig:finres} using an energy threshold of 50 EeV.}
\label{fig:fin50}
\end{figure*}

The entire analysis proposed in this paper, using cosmic rays above $10^{18.5}$~eV, has been repeated using a subset of cosmic rays with energies above 50~EeV. In this energy range $\sim$300 cosmic rays have been detected by Auger. Roughly this energy threshold is often used to search for correlations between cosmic rays and neutrinos, since it is commonly believed that at this energy the deflection for cosmic rays is small and correlations with neutrinos are more likely observed. However, at these very high energies the observed UHECRs come only from the very local Universe, while neutrinos can come also from distant sources (see upper panels of Fig.~\ref{plot:nuevo}). As a consequence, although at higher energies cosmic rays are less deflected, the cosmological connection with neutrinos is lost. Repeating the analysis with cosmic rays above 50~EeV, we obtain the results reported in Fig.~\ref{fig:fin50}. In order to observe correlations the local density should be much smaller than in the scenario with $E_\text{CR}>10^{18.5}$~eV (Fig.~\ref{fig:finres}); this suggests that $10^{18.5}$~eV would be a more appropriate cosmic-ray energy threshold for neutrino-UHECR correlation searches.

\bsp	
\label{lastpage}
\end{document}